# A BAT-based Exact-Solution Algorithm for the Series-Parallel Redundancy Allocation Problem with Mixed Components

Wei-Chang Yeh
Integration & Collaboration Laboratory
Department of Industrial Engineering and Engineering Management
National Tsing Hua University
yeh@ieee.org

**Abstract** -- The series-parallel (active) redundancy allocation problem with mixed components (RAP) involves setting reliable objectives for components or subsystems to meet the resource consumption constraint, e.g., the total cost. RAP has been an active research area for the past four decades. The NP-hard difficulties confronted by RAP are maintaining feasibility with respect to two constraints: cost and weight. A novel algorithm called the bound-rule-BAT (BRB) based on the binary-addition-tree algorithm (BAT), the dominance rule, and dynamic bounds are proposed to solve the exact solutions of the most famous RAP benchmark problems called the (33-variation) Fyffe RAP. From the experiments, the proposed BRB can solve the Fyffe RAP correctly under the assumption that the maximal number of components of each subsystem is eight, and this is the first exact-solution algorithm that can solve the Fyffe RAP within 8 seconds and 60 seconds if no reliability lower bound is used.

**Acronyms:**

    RAP: Redundancy allocation problem

    BAT: Binary-addition-tree algorithm

    BRB: Bound-Rule-BAT

**Notation:**

|•|: the number of elements in set •

$X$: the solution denoted the system structure for the RAP

$R(•)$: the reliability after given •

$W(•)$: the weight after given •

$C(•)$: the budget (cost) after given •

$W_{UB}$: the given limitation of the weight

$C_{UB}$: the given limitation of the cost

$r_{i,j}$: the reliability of the $j$th type of components in the $i$th subsystem

$c_{i,j}$: the cost of the $j$th type of components in the $i$th subsystem

$w_{i,j}$: the weight of the $j$th type of components in the $i$th subsystem

$R_{LB}$: the limitation of the reliability obtained from the other method

$n$: the total number of subsystems

$m_i$: the total number of different component types at the $i$th subsystems

$n_i$: the total number of redundancies at the $i$th subsystems

$B_i$: the ordered set of all $i$-tuple state vectors (without duplicates) obtained from the BAT

$X_{i,j,k}$: the $k$th coordinate of the $j$th vector in $B_i$

$B_{i,j}$: the super sub-BAT of $B_i$ such that $X_{i,l,k} = X_{j,l,k}$ for all $l, k = 1, 2, …, j$ and $j = 1, 2, …, i$.

$\otimes$: the multiplication operator in this study

$\Omega$: $\Omega = \{1, 2, …., 14\}$ is the set of subsystems in the Fyffe RAP

$\Omega_3$: $\Omega_3 = \{2, 4, 5, 7, 8, 10, 11, 13\}$ is the set of subsystems with 3 different types of components in the Fyffe RAP

$\Omega_4$: $\Omega_4 = \Omega - \Omega_3 = \{1, 3, 6, 9, 12, 14\}$ is the set of subsystems with 4 different types of components in the Fyffe RAP

$S_i$: $S_i = B_3$ if $i \in \Omega_3$ and $S_i = B_4$ if $i \in \Omega_4$

$C_n^m$: $C_n^m = \dfrac{m!}{n!(m-n)!}$

$Min(\bullet)$: the element with the minimum in set $\bullet$

$w_k$: $w_k = Min\{W(X) \mid \text{for all } X \in S_k\}$

$c_k$: $c_k = Min\{C(X) \mid \text{for all } X \in S_k\}$

$W_k$: $W_k = \sum_{i=k+1}^{n} w_i$

$C_k$: $C_k = \sum_{i=k+1}^{n} c_i$

$D(S_i)$: the new $S_i$ after implementing the proposed dominance rule

## 1. INTRODUCTION

The (active) redundancy allocation problem (RAP) is designed to maximize the system reliability and satisfy preset required conditions, e.g., the minimum cost and weight of manufacturing using redundant components in parallel [1, 2, 3]. RAP is the most significant problem in design-for-reliability and has been applied widely in the designing phase of numerous valuable engineering applications [4], industrial applications [5], and scientific applications [6] among different real-life reliability problems [7, 8] during the past thirty years.

RAP is also a well-known NP-hard problem, and its computational effort is increasing exponentially with the number of nodes and arcs in the system [9]. Hence the main focus has been on developing approximation methods, such as the Improved Surrogate Constraint Method [10, 11], Genetic Algorithm (GA) [12], Linear Programming Approach [13], Tabu Search [14], Ant Colony Optimization [15], Simulated Annealing Method [16, 17], the variable neighborhood search algorithm [18], Artificial Bee Colony [19], and Simplified

Swarm Optimization (SSO) [20] to solve RAP to avoid numerical difficulties and reduce computational burdens. Note that SSO is one of the best approximation methods for the problem currently.

To evaluate the quality and performance of the approximated methods, they are applied to the most cited benchmark problem (call the Fyffe RAP hereafter) in RAP, initially proposed by Fyffe *et al.* [1] and revised by Nakagawa [10]. This benchmark RAP is a set of 33-variation 14-subsystem of the original Fyffe problem.

The approximation methods can offer a tactical way of finding optimal or good quality solutions to larger problems. However, the exact solutions to the Fyffe RAP (more than 50 years ago) are still unknown if letting each subsystem have at most eight components, that is, we do not know how good are these results obtained from the approximated methods which all assume that each subsystem has at most eight components. Note that these exact methods have no clear information about the maximal number of components used in each subsystem [21, 22, 23].

Furthermore, the minimal/maximal number of components and the number of variables used in RAP that need to be used in this benchmark is also not known [21, 22, 23]. Thus, a method needs to be developed to find the exact solutions for Fyffe RAP to guide other approximation methods in finding optimal or good quality solutions to larger problems.

This study aims to develop the first algorithm called the BRB to obtain the exact solutions of the Fyffe RAP within 8 seconds or 60 seconds if without using the reliability lower bound. The BRB is based on the binary-state-tree algorithm (BAT) proposed by Yeh in [24]. The BAT has proven efficient and simple in finding all vectors and has been implemented in various versions to solve different problems [24, 25, 26]. The BRB also integrated the proposed dynamic bounds and the proposed dominance rule to reduce the solution space size to solve the Fyffe.

The rest of the paper is organized as follows. The overview of the RAP and BAT is provided in Section 2. Section 3 discusses the solution design adapted in the proposed BRB to improve its efficiency. The major components, including the multiplication of super sub-BATs, dynamic bounds, and the proposed dominance rule of the proposed BRB, are given in Section 4, together with its complete pseudocode. Section 5 illustrates the experiment results, including the solutions and the run times for the proposed BRB in solving the Fyffe RAP, the most famous RAP benchmarks with 33 variations. Final conclusions and future works are summarized in Section 6.

## 2. OVERVIEW OF RAP AND BAT

Section 2 briefly describes the RAP, the most famous RAP benchmark called the Fyffe RAP, and the traditional BAT, which is the basis of the proposed BRB.

### 2.1 RAP and Fyffe RAP

RAP with mixed components allows a subsystem to be duplicated with different sets of components [1, 10] and its formulation can be expressed as an integer nonlinear programming problem [10] as the following:

$$\text{Maximize} \quad R(X) \tag{1}$$

$$\text{s.t.} \quad C(X) \leq C_{UB} \tag{2}$$

$$W(X) \leq W_{UB}. \tag{3}$$

The objective nonlinear function Equation (1) is to maximize the series-parallel system reliability $R(X)$. Linear equations (2) and (3) limit the total cost $C(X)$ and weight $W(X)$ to less than or equal to predefined allowable amounts $C_{UB}$ and $W_{UB}$, respectively.

In the Fyffe RAP, there are $n = 14$ subsystems connected in series, and each subsystem can have at most eight components connected in parallel. There are 33 variations of the Fyffe

RAP with $C_{UB} = 130$ and $W_{UB} = 159, 160, \ldots, 191$. The corresponding values of $r_{i,j}$, $c_{i,j}$, and $w_{i,j}$ in the Fyffe RAP are given in Table 1.

Table 1. Data for the 33-variation benchmark problem taken from Fyffe [1, 10].

| | $r_{i,j}$ | | | | $c_{i,j}$ | | | | $w_{i,j}$ | | | | $i$ | $j$ | $r_{i,j}$ | | | | $c_{i,j}$ | | | | $w_{i,j}$ | | | |
|---|---|---|---|---|---|---|---|---|---|---|---|---|---|---|---|---|---|---|---|---|---|---|---|---|---|---|
| $i$ \ $j$ | 1 | 2 | 3 | 4 | 1 | 2 | 3 | 4 | 1 | 2 | 3 | 4 | | | 1 | 2 | 3 | 4 | 1 | 2 | 3 | 4 | 1 | 2 | 3 | 4 |
| 1 | .90 | .93 | .91 | .95 | 1 | 1 | 2 | 2 | 3 | 4 | 2 | 5 | 8 | | .81 | .90 | .91 | | 3 | 5 | 6 | | 4 | 7 | 6 | |
| 2 | .95 | .94 | .93 | | 2 | 1 | 1 | | 8 | 10 | 9 | | 9 | | .97 | .99 | .96 | .91 | 2 | 3 | 4 | 3 | 8 | 9 | 7 | 8 |
| 3 | .85 | .90 | .87 | .92 | 2 | 3 | 1 | 4 | 7 | 5 | 6 | 4 | 10 | | .83 | .85 | .90 | | 4 | 4 | 5 | | 6 | 5 | 6 | |
| 4 | .83 | .87 | .85 | | 3 | 4 | 5 | | 5 | 6 | 4 | | 11 | | .94 | .95 | .96 | | 3 | 4 | 5 | | 5 | 6 | 6 | |
| 5 | .94 | .93 | .95 | | 2 | 2 | 3 | | 4 | 3 | 5 | | 12 | | .79 | .82 | .85 | .90 | 2 | 3 | 4 | 5 | 4 | 5 | 6 | 7 |
| 6 | .99 | .98 | .97 | .96 | 3 | 3 | 2 | 2 | 5 | 4 | 5 | 4 | 13 | | .98 | .99 | .97 | | 2 | 3 | 2 | | 5 | 5 | 6 | |
| 7 | .91 | .92 | .94 | | 4 | 4 | 5 | | 7 | 8 | 9 | | 14 | | .90 | .92 | .95 | .99 | 4 | 4 | 5 | 6 | 6 | 7 | 6 | 9 |

The Fyffe RAP can be constructed as the series-parallel network shown in Figure 1 [20], where all components are parallel in each subsystem and all subsystems are connected in series and the number, say $i$, in each block denotes component $i$ used.

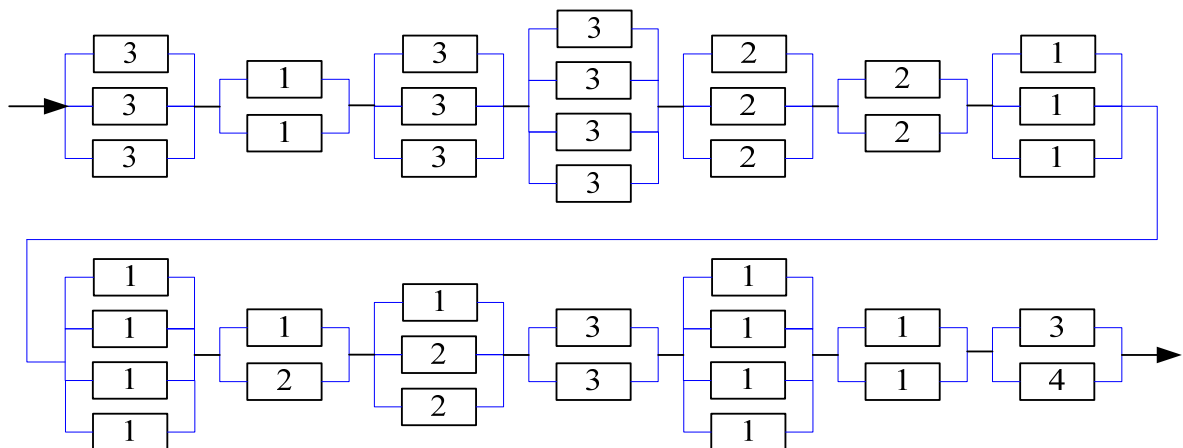

**Figure 1** Example RAP [20].

## 2.2 BAT

The basic concept of all BATs to list all binary-state vectors is simply by a procedure analogous to adding one to a binary code. After the emergence of the BAT proposed by Yeh in [24], there are various BATs proposed for different applications, that is, the backward BAT [24], the forward BAT [26, 28], the multi-state BAT [25], the node-based BAT [27], and so on [28-41].

The (forward) BAT revised in the proposed BRB adds one to the first coordinate and gradually moves to the last coordinate of the binary-state vectors, and its source code can be downloaded from [29]. The forward arc-based BAT [26, 28] is provided below.

**Input:** $\mu$.

**Output:** All $\mu$-tuple binary-state vectors.

**Algorithm: the (forward) BAT**

**STEP F0.** Let index $i$ and binary-state vector $X = (x_1, x_2, \ldots, x_\mu)$ be zero and vector zero, respectively.

**STEP F1.** If $x_i = 0$, let $x_i = i = 1$, a new $X$ is found, and go to STEP F1.

**STEP F2.** If $i = \mu$, halt and all binary-state vectors from vector zero to vector one are found.

**STEP F3.** Let $x_i = 0$, $i = i + 1$, and go to STEP F1.

STEP F0 initializes the first coordinate and the first binary-state vector to be $i = 0$ and $X = \mathbf{0}$. The loop from STEPs F1 to F3 creates all binary-state vectors analogous to the binary addition operator. If the current coordinate value is zero, that is, $x_i = 0$, $x_i$ is reset to one and a new $X$ is found in STEP F1. For example, the next $X = (1, 0, 1)$ if the current $X = (0, 0, 1)$.

STEP F2 stops the BAT if the index of the current coordinate is $\mu$ which implies that each $x_k$ is changed from one to zero for $k = 1, 2, \ldots, (\mu–1)$ already, that is, $X$ is already the vector one. For example, the next vector after $X = (1, 1, 1)$ is $(0, 0, 0, 1)$. Otherwise, as stated in STEP F3, the value of the current coordinate changes from one to zero and goes to STEP F1. For example, $x_1 = 1$ in $X = (1, 0, 1)$ is changed to $x_1 = 0$ and go to STEP F1.

The basic BAT has a time complexity $(2^{\mu+1})$ [25]. There are only four steps in BAT, and it is simple to code. We can add and calculate the related function values of $X$ for different-purpose problems in STEP F1 after obtaining each $X$, that is, we need to calculate $W(X)$, $C(X)$, and $R(X)$ in the RAP. Hence, the BAT is easy to make-to-fit. Moreover, from experiments,

BAT is more efficient than other search methods such as DFS and BFS. Hence, there are different BATs for different problems.

For example, Table 2 lists all 5-tuple binary-states obtained from the BAT using the above pseudocode.

**Table 2.** All vectors obtained from the binary-state BAT.

| $i$ | $X_i$ | $i$ | $X_i$ |
|---|---|---|---|
| 1 | (0, 0, 0, 0) | 9 | (0, 0, 0, 1) |
| 2 | (1, 0, 0, 0) | 10 | (1, 0, 0, 1) |
| 3 | (0, 1, 0, 0) | 11 | (0, 1, 0, 1) |
| 4 | (1, 1, 0, 0) | 12 | (1, 1, 0, 1) |
| 5 | (0, 0, 1, 0) | 13 | (0, 0, 1, 1) |
| 6 | (1, 0, 1, 0) | 14 | (1, 0, 1, 1) |
| 7 | (0, 1, 1, 0) | 15 | (0, 1, 1, 1) |
| 8 | (1, 1, 1, 0) | 16 | (1, 1, 1, 1) |

## 3. SOLUTION DESIGN

Each solution represents a system structure, and its design affects the size of the solution space and the efficiency of algorithms in solving the Fyffe RAP.

Two different solution designs represent 14 subsystems, with each subsystem including at least one redundancy and at most eight redundancies in Fyffe RAP. The first one is component-based, and the second is number-based. The details of these two different designs and the sizes of their solution spaces are discussed in this section separately.

### 3.1 Component-Based Solution Design

In the component-based solution design, each variable denotes the related component used in the system. Let $X = (y_{1,1}, y_{1,2}, \ldots, y_{1,n_1}, y_{2,1}, y_{2,2}, \ldots, y_{2,n_2}, \ldots, y_{n,1}, y_{n,2}, \ldots, y_{n,n_n})$ be a component-based solution and $y_{i,j}$ be the type of components used in the $j$th in the $i$th subsystems for all $i \in \Omega$ and $j = 1, 2, \ldots, n_i$. Note that $n_i = 8$ for all $i \in \Omega$ in the Fyffe RAP. Without loss of generality, we assume $y_{i,j} = 0$ means that there is no redundant component in the corresponding position. Hence, in the component-based solution design for the Fyffe RAP, we have

$$R(X) = \prod_{i=1}^{14}\left[1 - \prod_{j=1}^{8}(1 - R(y_{i,j}))\right] \tag{4}$$

$$C(X) = \sum_{i=1}^{14}\sum_{j=1}^{8} C(y_{i,j}) \tag{5}$$

$$W(X) = \sum_{i=1}^{14}\sum_{j=1}^{8} W(y_{i,j}) \tag{6}$$

$$y_{i,j} = \begin{cases} 0,1,2,3 & \text{if } i \in \Omega_3 \\ 0,1,2,3,4 & \text{if } i \in \Omega_3 \end{cases}. \tag{7}$$

For example, $X$ = (33300000, 11000000, 33300000, 33330000, 22200000, 22000000, 11100000, 1111 0000, 12000000, 12200000, 33000000, 11110000, 11000000, 34000000) is a component-based solution design [20] and its structure is depicted in Figure 1.

There are 14 subsystems, and each subsystem has 8 variables, that is, 14 × 8 = 112 variables in total. Because $m_i$ = 3 for $i \in \Omega_3$ and $m_i$ = 4 for $i \in \Omega_4$, we have $y_{i,j}$ = 0, 1, 2, 3 for $i \in \Omega_3$ and $y_{i,j}$ = 0, 1, 2, 3, 4 for $i \in \Omega_4$. Hence, there are $4^8$ and $5^8$ combinations for $i \in \Omega_3$ and $\Omega_4$, respectively, that is, the size of solution space is

$$(4^8)^8 \times (5^8)^6 = 1.20893\text{E}+72 \tag{8}$$

without considering the weights and costs.

## 3.2 Number-Based Solution Design

Let $X = (x_{1,1}, x_{1,2}, \ldots, x_{1,m_1}, x_{2,1}, x_{2,2}, \ldots, x_{2,m_2}, \ldots, x_{n,1}, x_{n,2}, \ldots, x_{n,m_n})$ be a number-based solution, where $x_{i,j}$ is the number of using the $j$th component of the $i$th subsystem for $i$ = 1, 2, ..., 14 and $j$ = 1, 2, ..., $m_i$, $m_i$ = 3 for $i \in \Omega_3$, and $m_i$ = 4 for $i \in \Omega_4$. Hence, there are 3 × $|\Omega_3|$ + 4 × $|\Omega_4|$ = 40 variables and

$$R(X) = \prod_{i=1}^{14}\left[1 - \prod_{j=1}^{m_i}(1 - r_{i,j})^{x_{i,j}}\right] \tag{9}$$

$$C(X) = \sum_{i=1}^{14} \sum_{j=1}^{m_i} x_{i,j} c_{i,j} \tag{10}$$

$$W(X) = \sum_{i=1}^{14} \sum_{j=1}^{m_i} x_{i,j} w_{i,j} \tag{11}$$

$$x_{i,j} = 0, 1, \ldots, 8. \tag{12}$$

For example, $X = (0030, 200, 0030, 004, 030, 0200, 300, 400, 1100, 120, 002, 4000, 200, 0011)$ is the number-based solution design for the system structure depicted in Figure 1.

Because

$$1 \leq \sum_{j=0}^{m_i} x_{i,j} \leq 8 \text{ and } x_{i,j} = 0, 1, 2, \ldots, 8, \tag{13}$$

For example, $X = (0030, 200, 0030, 004, 030, 0200, 300, 400, 1100, 120, 002, 4000, 200, 0011)$ is the number-based solution design for the system structure depicted in Figure 1.

Because

$$1 \leq \sum_{j=0}^{m_i} x_{i,j} \leq 8 \text{ and } x_{i,j} = 0, 1, 2, \ldots, 8, \tag{13}$$

the total number of combinations of different $x_{i,j}$ for $i \in \Omega_3$ and for $i \in \Omega_4$ are

$$C_2^{1+2} + C_2^{2+2} + C_2^{3+2} + C_2^{4+2} + C_2^{5+2} + C_2^{6+2} + C_2^{7+2} + C_2^{8+2} = 164 \tag{14}$$

and

$$C_3^{1+3} + C_3^{2+3} + C_3^{3+3} + C_3^{4+3} + C_3^{5+3} + C_3^{6+3} + C_3^{7+3} + C_3^{8+3} = 494, \tag{15}$$

respectively. Thus, without considering the weights and costs, the size of the solution space based on the number-based solution design is

$$(164)^8 \times (494)^6 = 7.60523\text{E}+33. \tag{16}$$

From Eq. (8) and (16), we have 7.60523E+33<<1.20893E+72, that is, the solution space of the number-based solution expression is smaller than that of the component-based solution design. Hence, the number-based solution design is implemented in the proposed BRB.

## 4. PROPOSED BRB FOR RAP

The proposed BRB is based on the proposed upper-bound BAT, the weight and cost dynamic bound, reliability bound $R_{LB}$, and the dominance rule to find the exact reliability of the 33-variation benchmark problem taken from Fyffe. This section presents the overall process for the proposed BRB for solving the exact reliability of the Fyffe RAP.

### 4.1 Proposed Upper-Bound BAT

A new BAT called the upper-bound BAT is proposed to find all possible vectors in $B_u$ such that the value of each coordinate is a nonnegative integer, and the summation of all coordinate values is less than u.

**Algorithm: Upper-Bound BAT**

**Input:** $\mu$ and u > 1.

**Output:** All $\mu$-tuple state vectors $X = (x_1, x_2, \ldots, x_\mu)$ such that $\sum_{i=1}^{\mu} x_i < u$ and $x_i = 0, 1, \ldots,$ (u−1) for $i = 1, 2, \ldots, \mu$.

**STEP B0.** Let $i$ = SUM = 0 and $X = \mathbf{0}$.

**STEP B1.** If SUM < (u−1), let $x_i = x_i + 1$, SUM = SUM + 1, a new $X$ is found, and go to STEP B1.

**STEP B2.** If $i = \mu$, halt and all related vectors are found.

**STEP B3.** Let SUM = SUM − $x_i$, $x_i = 0$, $i = i + 1$, and go to STEP B1.

For example, the first 210 vectors and related information in $B_4$ are listed in Table 3, where notations $S = (x_1 + x_2 + x_3 + x_4) = 1, 2, \ldots, 8$, $W = (w_{1,1}x_1 + w_{1,2}x_2 + w_{1,3}x_3 + w_{1,4}x_4)$, $C = (c_{1,1}x_1 + c_{1,2}x_2 + c_{1,3}x_3 + c_{1,4}x_4)$, and $R = 1 - (1-r_{1,1})^{x_1}(1-r_{1,2})^{x_2}(1-r_{1,3})^{x_3}(1-r_{1,4})^{x_4}$, and $w_{1,i}$, $c_{1,i}$ and $r_{1,i}$ are listed in Table 1 for $i = 1, 2, 3, 4$.

Note that $(x_1, x_2, x_3, x_4) = (0, 0, 0, 0)$ is removed because there is at least one component in each subsystem and each vector in $B_4$ represents a subsystem with at most four different types of components.

**Table 3.** The first 210 vectors and related information in $B_4$.

| k | $x_1$ | $x_2$ | $x_3$ | $x_4$ | S | W | C | R | k | $x_1$ | $x_2$ | $x_3$ | $x_4$ | S | W | C | R | k | $x_1$ | $x_2$ | $x_3$ | $x_4$ | S | W | C | R |
|---|---|---|---|---|---|---|---|---|---|---|---|---|---|---|---|---|---|---|---|---|---|---|---|---|---|---|
| 1 | 1 | 0 | 0 | 0 | 1 | 3 | 1 | 0.900000 | 71 | 0 | 4 | 1 | 0 | 5 | 18 | 6 | 0.999998 | 141 | 2 | 2 | 4 | 0 | 8 | 22 | 12 | 1.000000 |
| 2 | 2 | 0 | 0 | 0 | 2 | 6 | 2 | 0.990000 | 72 | 1 | 4 | 1 | 0 | 6 | 21 | 7 | 1.000000 | 142 | 0 | 3 | 4 | 0 | 7 | 20 | 11 | 1.000000 |
| 3 | 3 | 0 | 0 | 0 | 3 | 9 | 3 | 0.999000 | 73 | 2 | 4 | 1 | 0 | 7 | 24 | 8 | 1.000000 | 143 | 1 | 3 | 4 | 0 | 8 | 23 | 12 | 1.000000 |
| 4 | 4 | 0 | 0 | 0 | 4 | 12 | 4 | 0.999900 | 74 | 3 | 4 | 1 | 0 | 8 | 27 | 9 | 1.000000 | 144 | 0 | 4 | 4 | 0 | 8 | 24 | 12 | 1.000000 |
| 5 | 5 | 0 | 0 | 0 | 5 | 15 | 5 | 0.999990 | 75 | 0 | 5 | 1 | 0 | 6 | 22 | 7 | 1.000000 | 145 | 0 | 0 | 5 | 0 | 5 | 10 | 10 | 0.999994 |
| 6 | 6 | 0 | 0 | 0 | 6 | 18 | 6 | 0.999999 | 76 | 1 | 5 | 1 | 0 | 7 | 25 | 8 | 1.000000 | 146 | 1 | 0 | 5 | 0 | 6 | 13 | 11 | 0.999999 |
| 7 | 7 | 0 | 0 | 0 | 7 | 21 | 7 | 1.000000 | 77 | 2 | 5 | 1 | 0 | 8 | 28 | 9 | 1.000000 | 147 | 2 | 0 | 5 | 0 | 7 | 16 | 12 | 1.000000 |
| 8 | 8 | 0 | 0 | 0 | 8 | 24 | 8 | 1.000000 | 78 | 0 | 6 | 1 | 0 | 7 | 26 | 8 | 1.000000 | 148 | 3 | 0 | 5 | 0 | 8 | 19 | 13 | 1.000000 |
| 9 | 0 | 1 | 0 | 0 | 1 | 4 | 1 | 0.930000 | 79 | 1 | 6 | 1 | 0 | 8 | 29 | 9 | 1.000000 | 149 | 0 | 1 | 5 | 0 | 6 | 14 | 11 | 1.000000 |
| 10 | 1 | 1 | 0 | 0 | 2 | 7 | 2 | 0.993000 | 80 | 0 | 7 | 1 | 0 | 8 | 30 | 9 | 1.000000 | 150 | 1 | 1 | 5 | 0 | 7 | 17 | 12 | 1.000000 |
| 11 | 2 | 1 | 0 | 0 | 3 | 10 | 3 | 0.999300 | 81 | 0 | 0 | 2 | 0 | 2 | 4 | 4 | 0.991900 | 151 | 2 | 1 | 5 | 0 | 8 | 20 | 13 | 1.000000 |
| 12 | 3 | 1 | 0 | 0 | 4 | 13 | 4 | 0.999930 | 82 | 1 | 0 | 2 | 0 | 3 | 7 | 5 | 0.999190 | 152 | 0 | 2 | 5 | 0 | 7 | 18 | 12 | 1.000000 |
| 13 | 4 | 1 | 0 | 0 | 5 | 16 | 5 | 0.999993 | 83 | 2 | 0 | 2 | 0 | 4 | 10 | 6 | 0.999919 | 153 | 1 | 2 | 5 | 0 | 8 | 21 | 13 | 1.000000 |
| 14 | 5 | 1 | 0 | 0 | 6 | 19 | 6 | 0.999999 | 84 | 3 | 0 | 2 | 0 | 5 | 13 | 7 | 0.999992 | 154 | 0 | 3 | 5 | 0 | 8 | 22 | 13 | 1.000000 |
| 15 | 6 | 1 | 0 | 0 | 7 | 22 | 7 | 1.000000 | 85 | 4 | 0 | 2 | 0 | 6 | 16 | 8 | 0.999999 | 155 | 0 | 0 | 6 | 0 | 6 | 12 | 12 | 0.999999 |
| 16 | 7 | 1 | 0 | 0 | 8 | 25 | 8 | 1.000000 | 86 | 5 | 0 | 2 | 0 | 7 | 19 | 9 | 1.000000 | 156 | 1 | 0 | 6 | 0 | 7 | 15 | 13 | 1.000000 |
| 17 | 0 | 2 | 0 | 0 | 2 | 8 | 2 | 0.995100 | 87 | 6 | 0 | 2 | 0 | 8 | 22 | 10 | 1.000000 | 157 | 2 | 0 | 6 | 0 | 8 | 18 | 14 | 1.000000 |
| 18 | 1 | 2 | 0 | 0 | 3 | 11 | 3 | 0.999510 | 88 | 0 | 1 | 2 | 0 | 3 | 8 | 5 | 0.999433 | 158 | 0 | 1 | 6 | 0 | 7 | 16 | 13 | 1.000000 |
| 19 | 2 | 2 | 0 | 0 | 4 | 14 | 4 | 0.999951 | 89 | 1 | 1 | 2 | 0 | 4 | 11 | 6 | 0.999943 | 159 | 1 | 1 | 6 | 0 | 8 | 19 | 14 | 1.000000 |
| 20 | 3 | 2 | 0 | 0 | 5 | 17 | 5 | 0.999995 | 90 | 2 | 1 | 2 | 0 | 5 | 14 | 7 | 0.999994 | 160 | 0 | 2 | 6 | 0 | 8 | 20 | 14 | 1.000000 |
| 21 | 4 | 2 | 0 | 0 | 6 | 20 | 6 | 1.000000 | 91 | 3 | 1 | 2 | 0 | 6 | 17 | 8 | 0.999999 | 161 | 0 | 0 | 7 | 0 | 7 | 14 | 14 | 1.000000 |
| 22 | 5 | 2 | 0 | 0 | 7 | 23 | 7 | 1.000000 | 92 | 4 | 1 | 2 | 0 | 7 | 20 | 9 | 1.000000 | 162 | 1 | 0 | 7 | 0 | 8 | 17 | 15 | 1.000000 |
| 23 | 6 | 2 | 0 | 0 | 8 | 26 | 8 | 1.000000 | 93 | 5 | 1 | 2 | 0 | 8 | 23 | 10 | 1.000000 | 163 | 0 | 1 | 7 | 0 | 8 | 18 | 15 | 1.000000 |
| 24 | 0 | 3 | 0 | 0 | 3 | 12 | 3 | 0.999657 | 94 | 0 | 2 | 2 | 0 | 4 | 12 | 6 | 0.999960 | 164 | 0 | 0 | 8 | 0 | 8 | 16 | 16 | 1.000000 |
| 25 | 1 | 3 | 0 | 0 | 4 | 15 | 4 | 0.999966 | 95 | 1 | 2 | 2 | 0 | 5 | 15 | 7 | 0.999996 | 165 | 0 | 0 | 0 | 1 | 1 | 5 | 2 | 0.950000 |
| 26 | 2 | 3 | 0 | 0 | 5 | 18 | 5 | 0.999997 | 96 | 2 | 2 | 2 | 0 | 6 | 18 | 8 | 1.000000 | 166 | 1 | 0 | 0 | 1 | 2 | 8 | 3 | 0.995000 |
| 27 | 3 | 3 | 0 | 0 | 6 | 21 | 6 | 1.000000 | 97 | 3 | 2 | 2 | 0 | 7 | 21 | 9 | 1.000000 | 167 | 2 | 0 | 0 | 1 | 3 | 11 | 4 | 0.999500 |
| 28 | 4 | 3 | 0 | 0 | 7 | 24 | 7 | 1.000000 | 98 | 4 | 2 | 2 | 0 | 8 | 24 | 10 | 1.000000 | 168 | 3 | 0 | 0 | 1 | 4 | 14 | 5 | 0.999950 |
| 29 | 5 | 3 | 0 | 0 | 8 | 27 | 8 | 1.000000 | 99 | 0 | 3 | 2 | 0 | 5 | 16 | 7 | 0.999997 | 169 | 4 | 0 | 0 | 1 | 5 | 17 | 6 | 0.999995 |
| 30 | 0 | 4 | 0 | 0 | 4 | 16 | 4 | 0.999976 | 100 | 1 | 3 | 2 | 0 | 6 | 19 | 8 | 1.000000 | 170 | 5 | 0 | 0 | 1 | 6 | 20 | 7 | 1.000000 |
| 31 | 1 | 4 | 0 | 0 | 5 | 19 | 5 | 0.999998 | 101 | 2 | 3 | 2 | 0 | 7 | 22 | 9 | 1.000000 | 171 | 6 | 0 | 0 | 1 | 7 | 23 | 8 | 1.000000 |
| 32 | 2 | 4 | 0 | 0 | 6 | 22 | 6 | 1.000000 | 102 | 3 | 3 | 2 | 0 | 8 | 25 | 10 | 1.000000 | 172 | 7 | 0 | 0 | 1 | 8 | 26 | 9 | 1.000000 |
| 33 | 3 | 4 | 0 | 0 | 7 | 25 | 7 | 1.000000 | 103 | 0 | 4 | 2 | 0 | 6 | 20 | 8 | 1.000000 | 173 | 0 | 1 | 0 | 1 | 2 | 9 | 3 | 0.996500 |
| 34 | 4 | 4 | 0 | 0 | 8 | 28 | 8 | 1.000000 | 104 | 1 | 4 | 2 | 0 | 7 | 23 | 9 | 1.000000 | 174 | 1 | 1 | 0 | 1 | 3 | 12 | 4 | 0.999650 |
| 35 | 0 | 5 | 0 | 0 | 5 | 20 | 5 | 0.999998 | 105 | 2 | 4 | 2 | 0 | 8 | 26 | 10 | 1.000000 | 175 | 2 | 1 | 0 | 1 | 4 | 15 | 5 | 0.999965 |
| 36 | 1 | 5 | 0 | 0 | 6 | 23 | 6 | 1.000000 | 106 | 0 | 5 | 2 | 0 | 7 | 24 | 9 | 1.000000 | 176 | 3 | 1 | 0 | 1 | 5 | 18 | 6 | 0.999997 |
| 37 | 2 | 5 | 0 | 0 | 7 | 26 | 7 | 1.000000 | 107 | 1 | 5 | 2 | 0 | 8 | 27 | 10 | 1.000000 | 177 | 4 | 1 | 0 | 1 | 6 | 21 | 7 | 1.000000 |
| 38 | 3 | 5 | 0 | 0 | 8 | 29 | 8 | 1.000000 | 108 | 0 | 6 | 2 | 0 | 8 | 28 | 10 | 1.000000 | 178 | 5 | 1 | 0 | 1 | 7 | 24 | 8 | 1.000000 |
| 39 | 0 | 6 | 0 | 0 | 6 | 24 | 6 | 1.000000 | 109 | 0 | 0 | 3 | 0 | 3 | 6 | 6 | 0.999271 | 179 | 6 | 1 | 0 | 1 | 8 | 27 | 9 | 1.000000 |
| 40 | 1 | 6 | 0 | 0 | 7 | 27 | 7 | 1.000000 | 110 | 1 | 0 | 3 | 0 | 4 | 9 | 7 | 0.999927 | 180 | 0 | 2 | 0 | 1 | 3 | 13 | 4 | 0.999755 |
| 41 | 2 | 6 | 0 | 0 | 8 | 30 | 8 | 1.000000 | 111 | 2 | 0 | 3 | 0 | 5 | 12 | 8 | 0.999993 | 181 | 1 | 2 | 0 | 1 | 4 | 16 | 5 | 0.999976 |
| 42 | 0 | 7 | 0 | 0 | 7 | 28 | 7 | 1.000000 | 112 | 3 | 0 | 3 | 0 | 6 | 15 | 9 | 0.999999 | 182 | 2 | 2 | 0 | 1 | 5 | 19 | 6 | 0.999998 |
| 43 | 1 | 7 | 0 | 0 | 8 | 31 | 8 | 1.000000 | 113 | 4 | 0 | 3 | 0 | 7 | 18 | 10 | 1.000000 | 183 | 3 | 2 | 0 | 1 | 6 | 22 | 7 | 1.000000 |
| 44 | 0 | 8 | 0 | 0 | 8 | 32 | 8 | 1.000000 | 114 | 5 | 0 | 3 | 0 | 8 | 21 | 11 | 1.000000 | 184 | 4 | 2 | 0 | 1 | 7 | 25 | 8 | 1.000000 |
| 45 | 0 | 0 | 1 | 0 | 1 | 2 | 2 | 0.910000 | 115 | 0 | 1 | 3 | 0 | 4 | 10 | 7 | 0.999949 | 185 | 5 | 2 | 0 | 1 | 8 | 28 | 9 | 1.000000 |
| 46 | 1 | 0 | 1 | 0 | 2 | 5 | 3 | 0.991000 | 116 | 1 | 1 | 3 | 0 | 5 | 13 | 8 | 0.999995 | 186 | 0 | 3 | 0 | 1 | 4 | 17 | 5 | 0.999983 |
| 47 | 2 | 0 | 1 | 0 | 3 | 8 | 4 | 0.999100 | 117 | 2 | 1 | 3 | 0 | 6 | 16 | 9 | 0.999999 | 187 | 1 | 3 | 0 | 1 | 5 | 20 | 6 | 0.999998 |
| 48 | 3 | 0 | 1 | 0 | 4 | 11 | 5 | 0.999910 | 118 | 3 | 1 | 3 | 0 | 7 | 19 | 10 | 1.000000 | 188 | 2 | 3 | 0 | 1 | 6 | 23 | 7 | 1.000000 |
| 49 | 4 | 0 | 1 | 0 | 5 | 14 | 6 | 0.999991 | 119 | 4 | 1 | 3 | 0 | 8 | 22 | 11 | 1.000000 | 189 | 3 | 3 | 0 | 1 | 7 | 26 | 8 | 1.000000 |
| 50 | 5 | 0 | 1 | 0 | 6 | 17 | 7 | 0.999999 | 120 | 0 | 2 | 3 | 0 | 5 | 14 | 8 | 0.999996 | 190 | 4 | 3 | 0 | 1 | 8 | 29 | 9 | 1.000000 |
| 51 | 6 | 0 | 1 | 0 | 7 | 20 | 8 | 1.000000 | 121 | 1 | 2 | 3 | 0 | 6 | 17 | 9 | 1.000000 | 191 | 0 | 4 | 0 | 1 | 5 | 21 | 6 | 0.999999 |

| 52 | 7 | 0 | 1 | 0 | 8 | 23 | 9 | 1.000000 | 122 | 2 | 2 | 3 | 0 | 7 | 20 | 10 | 1.000000 | 192 | 1 | 4 | 0 | 1 | 6 | 24 | 7 | 1.000000 |
|---|---|---|---|---|---|---|---|---|---|---|---|---|---|---|---|---|---|---|---|---|---|---|---|---|---|---|
| 53 | 0 | 1 | 1 | 0 | 2 | 6 | 3 | 0.993700 | 123 | 3 | 2 | 3 | 0 | 8 | 23 | 11 | 1.000000 | 193 | 2 | 4 | 0 | 1 | 7 | 27 | 8 | 1.000000 |
| 54 | 1 | 1 | 1 | 0 | 3 | 9 | 4 | 0.999370 | 124 | 0 | 3 | 3 | 0 | 6 | 18 | 9 | 1.000000 | 194 | 3 | 4 | 0 | 1 | 8 | 30 | 9 | 1.000000 |
| 55 | 2 | 1 | 1 | 0 | 4 | 12 | 5 | 0.999937 | 125 | 1 | 3 | 3 | 0 | 7 | 21 | 10 | 1.000000 | 195 | 0 | 5 | 0 | 1 | 6 | 25 | 7 | 1.000000 |
| 56 | 3 | 1 | 1 | 0 | 5 | 15 | 6 | 0.999994 | 126 | 2 | 3 | 3 | 0 | 8 | 24 | 11 | 1.000000 | 196 | 1 | 5 | 0 | 1 | 7 | 28 | 8 | 1.000000 |
| 57 | 4 | 1 | 1 | 0 | 6 | 18 | 7 | 0.999999 | 127 | 0 | 4 | 3 | 0 | 7 | 22 | 10 | 1.000000 | 197 | 2 | 5 | 0 | 1 | 8 | 31 | 9 | 1.000000 |
| 58 | 5 | 1 | 1 | 0 | 7 | 21 | 8 | 1.000000 | 128 | 1 | 4 | 3 | 0 | 8 | 25 | 11 | 1.000000 | 198 | 0 | 6 | 0 | 1 | 7 | 29 | 8 | 1.000000 |
| 59 | 6 | 1 | 1 | 0 | 8 | 24 | 9 | 1.000000 | 129 | 0 | 5 | 3 | 0 | 8 | 26 | 11 | 1.000000 | 199 | 1 | 6 | 0 | 1 | 8 | 32 | 9 | 1.000000 |
| 60 | 0 | 2 | 1 | 0 | 3 | 10 | 4 | 0.999559 | 130 | 0 | 0 | 4 | 0 | 4 | 8 | 8 | 0.999934 | 200 | 0 | 7 | 0 | 1 | 8 | 33 | 9 | 1.000000 |
| 61 | 1 | 2 | 1 | 0 | 4 | 13 | 5 | 0.999956 | 131 | 1 | 0 | 4 | 0 | 5 | 11 | 9 | 0.999993 | 201 | 0 | 0 | 1 | 1 | 2 | 7 | 4 | 0.995500 |
| 62 | 2 | 2 | 1 | 0 | 5 | 16 | 6 | 0.999996 | 132 | 2 | 0 | 4 | 0 | 6 | 14 | 10 | 0.999999 | 202 | 1 | 0 | 1 | 1 | 3 | 10 | 5 | 0.999550 |
| 63 | 3 | 2 | 1 | 0 | 6 | 19 | 7 | 1.000000 | 133 | 3 | 0 | 4 | 0 | 7 | 17 | 11 | 1.000000 | 203 | 2 | 0 | 1 | 1 | 4 | 13 | 6 | 0.999955 |
| 64 | 4 | 2 | 1 | 0 | 7 | 22 | 8 | 1.000000 | 134 | 4 | 0 | 4 | 0 | 8 | 20 | 12 | 1.000000 | 204 | 3 | 0 | 1 | 1 | 5 | 16 | 7 | 0.999996 |
| 65 | 5 | 2 | 1 | 0 | 8 | 25 | 9 | 1.000000 | 135 | 0 | 1 | 4 | 0 | 5 | 12 | 9 | 0.999995 | 205 | 4 | 0 | 1 | 1 | 6 | 19 | 8 | 1.000000 |
| 66 | 0 | 3 | 1 | 0 | 4 | 14 | 5 | 0.999969 | 136 | 1 | 1 | 4 | 0 | 6 | 15 | 10 | 1.000000 | 206 | 5 | 0 | 1 | 1 | 7 | 22 | 9 | 1.000000 |
| 67 | 1 | 3 | 1 | 0 | 5 | 17 | 6 | 0.999997 | 137 | 2 | 1 | 4 | 0 | 7 | 18 | 11 | 1.000000 | 207 | 6 | 0 | 1 | 1 | 8 | 25 | 10 | 1.000000 |
| 68 | 2 | 3 | 1 | 0 | 6 | 20 | 7 | 1.000000 | 138 | 3 | 1 | 4 | 0 | 8 | 21 | 12 | 1.000000 | 208 | 0 | 1 | 1 | 1 | 3 | 11 | 5 | 0.999685 |
| 69 | 3 | 3 | 1 | 0 | 7 | 23 | 8 | 1.000000 | 139 | 0 | 2 | 4 | 0 | 6 | 16 | 10 | 1.000000 | 209 | 1 | 1 | 1 | 1 | 4 | 14 | 6 | 0.999969 |
| 70 | 4 | 3 | 1 | 0 | 8 | 26 | 9 | 1.000000 | 140 | 1 | 2 | 4 | 0 | 7 | 19 | 11 | 1.000000 | 210 | 2 | 1 | 1 | 1 | 5 | 17 | 7 | 0.999997 |

### 4.2 Super Sub-BAT

A super sub-BAT $B_j$ is a special sub-BAT of $B_i$ such that $X \in B_j$ if and only if $X$ is a sub-vector with the first $j$ coordinates of a vector in $B_i$ [26].

The following lemma shows that we can simply find $B_i$ if we need to find any subsets of $\{B_1, B_2, …, B_{(i-1)}\}$ and each $B_j$ is a super sub-BAT of $B_i$ for $j = 1, 2, …, (i–1)$.

**Lemma 1.** $B_{i,j} = B_j$.

**Proof.**

The first vector of $B_i$ and $B_{i,j}$ are vector zero excepted that the former is $i$-tuple and the latter is $j$-tuple, that is, $X_{i,1,k} = X_{j,1,k} = 0$ for all $k \leq j$. Assume that $X_{i,h,k} = X_{j,h,k}$ for all $k \leq j$, $h = 1, 2, …, n$, and $n < |B_{i,j}|$. Consider the following three cases by induction:

**Case 1:** Let $\beta < \alpha \leq j$, $X_{j,n,\beta} < (u–1)$ and $X_{j,n,\alpha} = (u–1)$. From STEPs F1 in Section 2.2 or B1 in Section 4.1, $X_{j,n+1,\beta} = X_{j,n,\beta}$ for all $\beta \neq 1$ and $X_{j,n+1,1} = X_{j,n+1,1} + 1$. In the same way, we have $X_{i,n+1,\beta} = X_{i,n,\beta}$ for all $\beta \neq 1$ and $X_{i,n+1,1} = X_{i,n+1,1} + 1$

**Case 2:** Let $\beta < \alpha \leq j$, $X_{j,n,\beta} = (u–1)$, and $X_{j,n,\alpha} < (u–1)$. From STEPs F3 in Section 2.2 or B3 in Section 4.1, $X_{j,n+1,\beta} = 0$ and $X_{j,n+1,\alpha} = X_{j,n,\alpha} + 1$. In the same way, we have $X_{i,n+1,\beta} = 0$ and

$X_{i,n+1,\alpha} = X_{i,n,\alpha} + 1$.

**Case 3:** Let $X_{j,n,\beta} = (u-1)$ for all $\beta \leq j$. It is trivial that $X_{j,n}$ is the last state vector obtained from the BAT, i.e., $n = |B_{i,j}|$ which is contradict to that $n < |B_{i,j}|$.

From the above three cases, $X_{j,n+1,\alpha} = X_{i,n+1,\alpha}$ for $\alpha = 1, 2, \ldots, j$ if $n < |B_{i,j}|$. Hence, this lemma is correct from the induction method. □

For example, $B_1 = \{(0), (1)\} = B_{4,1}$, $B_2 = \{(0, 0), (1, 0), (0, 1), (1, 1)\} = B_{4,2}$, where $B_4$ is listed in Table 2.

Super sub-BATs can be multiplied in series using the operator $\otimes$ to have a large super sub-BAT which is called a convolutional sub-BAT as stated in the below lemma.

**Lemma 2.** $B_i \otimes B_j = B_{(i+j)}$.

**Proof.**

$B_{(i+j)}$, $B_i$, and $B_j$ include any feasible $(i+j)$-tuple, $i$-tuple, and $j$-tuple vectors, respectively.

□

Lemma 2 is a reverse concept of Lemma 1. The following corollary shows the relationship between $|B_i| \otimes |B_j|$ and $|B_{(i+j)}|$.

**Corollary 1.** $|B_i| \times |B_j| = |B_i| \otimes |B_j| = |B_i \otimes B_j|$.

For example, let $B_1 = \{(0), (1)\}$ and $B_2 = \{(0, 0), (1, 0), (0, 1), (1, 1)\}$. We have $B_1 \otimes B_2$ = $\{(0, 0, 0), (1, 0, 0), (0, 1, 0), (1, 1, 0), (0, 0, 1), (1, 0, 1), (0, 1, 1), (1, 1, 1)\} = B_{(1+2)} = B_3$ and $|B_1| \otimes |B_2| = 2^1 \times 2^2 = 2^3 = |B_1| \otimes |B_2| = |B_3|$.

**Corollary 2.** $X_{i,a} \otimes X_{j,b} = X_{(i+j),i(b-1)+a}$.

**Proof.**

Because $X_{i,a} \otimes X_{j,b} = (X_{i,a,1}, X_{i,a,2}, \ldots, X_{i,a,i}, X_{j,b,1}, X_{j,b,2}, \ldots, X_{j,b,j})$ and $B_i \otimes B_j = B_{(i+j)}$, $(X_{i,a,1}, X_{i,a,2}, \ldots, X_{i,a,i}, X_{j,b,1}, X_{j,b,2}, \ldots, X_{j,b,j}) = (X_{i+j,i(b-1)+a,1}, X_{i+j,i(b-1)+a,2}, \ldots, X_{i+j,i(b-1)+a,i+j})$. □

For example, $X_{1,2} = (1)$ and $X_{2,3} = (0, 1)$, and we have $X_{1,2} \otimes X_{2,3} = (1, 0, 1) = X_{3,6}$.

**Corollary 3.** $W(X_{i,a} \otimes X_{j,b}) = W(X_{i,a}) + W(X_{j,b})$, $C(X_{i,a} \otimes X_{j,b}) = C(X_{i,a}) + C(X_{j,b})$, $R(X_{i,a} \otimes X_{j,b}) = R(X_{i,a}) \times R(X_{j,b})$.

Let convolutional super sub-vector $\bigotimes_{i=1}^{k} X_i = X_1 \otimes X_2 \otimes \ldots \otimes X_k$, and convolutional super subsystem $\bigotimes_{i=1}^{k} S_i = S_1 \otimes S_2 \otimes \ldots \otimes S_k$, where $X_i \in S_i$ for all $i = 1, 2, \ldots, k$, $S_i = B_3$ if $i \in \Omega_3$ and $S_i = B_4$ if $i \in \Omega_4$. From Corollaries 2 and 3, we have the following lemma:

**Lemma 3.** $W(\bigotimes_{i=1}^{k} X_i) = \sum_{i=1}^{k} W(X_i)$, $C(\bigotimes_{i=1}^{k} X_i) = \sum_{i=1}^{k} C(X_i)$, and $R(\bigotimes_{i=1}^{k} X_i) = \prod_{i=1}^{k} R(X_i)$, where $X_i \in S_i$ for all $i = 1, 2, \ldots, k$.

For example, let $X_1 = (0, 0, 3, 0)$ and $X_2 = (2, 0, 0)$. Based on Table 1, we have

$$X_1 \otimes X_2 = (0, 0, 3, 0, 2, 0, 0) \tag{17}$$

$$C(X_1 \otimes X_2) = C(X_1) + C(X_2) = C(0, 0, 3, 0, 2, 0, 0) = 3 \times 2 + 2 \times 2 = 10 \tag{18}$$

$$W(X_1 \otimes X_2) = W(X_1) + W(X_2) = W(0, 0, 3, 0, 2, 0, 0) = 3 \times 2 + 2 \times 8 = 22 \tag{19}$$

$$R(X_1 \otimes X_2) = R(X_1) \times R(X_2) = R(0, 0, 3, 0, 2, 0, 0)$$

$$= [1-(1-.91)^3] \times [1-(1-.95)^2] = 0.996772823. \tag{20}$$

From the above discussion, we can have one prototype BAT with all 4-tuple binary-state vectors, that is, $B_4$, for the Fyffe RAP, and $B_3$ is a super sub-$B_4$ from Lemma 1. Hence, $|S_i| =$

$\sum_{k=1}^{8} C_k^{2+k} = 164$ and $\sum_{k=1}^{8} C_k^{3+k} = 494$ for $i \in \Omega_3$ and $\Omega_4$, respectively. Thus, all solutions for the 14-subsystem Fyffe RAP are included in $\bigotimes_{i=1}^{14} S_i = (S_1 \otimes S_2 \otimes \ldots \otimes S_{14})$ from Lemma 3.

### 4.3 $R_{LB}$

All subsystems are connected in series in the Fyffe RAP. Hence, we have the following lemma.

**Lemma 4.** $R(\bigotimes_{i=1}^{k} X_i) \leq \text{Min}\{R(X_i) \mid X_i \in S_i \text{ for } i = 1, 2, \ldots, k\}$.

**Proof.**

It is trivial because $R(\bigotimes_{i=1}^{k} X_i) = R(X_1) \times R(X_2) \times \ldots \times R(X_k)$ and $0 < R(X_i) \leq 1$ for $i = 1, 2, \ldots, k$. ☐

**Corollary 4.** $\bigotimes_{i=1}^{k} X_i$ is an infeasible super sub-vector, where $X_i \in S_i$ for $i = 1, 2\ldots, k$, if

$$R(\bigotimes_{i=1}^{k} X_i) < R_{LB}. \tag{23}$$

The original values of $|S_k|$, $w_k$, $c_k$, $W_k$, and $C_k$ without using $R_{LB}$, the proposed dominance rule, and the dynamic bounds are listed in these columns under "Original" in Table 4 for $k = 1, 2, \ldots, 14$.

These alphabets $R$, $N$, $W$, and $C$ in columns under "$RN$" and "$RNWC$" are represented $R_{LB}$, the proposed dominance rule, the dynamic weight bound $W_{LB}$, and the dynamic cost bound $C_{LB}$ are implemented, respectively. The proposed dominance rule, $W_{LB}$, and $C_{LB}$ are discussed in the next subsections.

These columns under "$R$" are the new values of $|S_k|$, $w_k$, $c_k$, $W_k$, and $C_k$ after considering

$R_{LB}$ = 0.954565 obtained from SSO proposed in [20] for $C_{UB}$ = 130 and $W_{UB}$ = 159. For example, $X_1$ = (1, 0, 0, 0) and $R(X_1)$ = 0.9 < 0.954565 in Table 4. Hence, $X_1$ can be removed.

Similarly, each value of $|S_k|$ in the column under "RN" and "RNWC" is the new value of $|S_k|$ after using ($R_{LB}$ and the proposed dominance rule), and ($R_{LB}$, the proposed dominance rule, the dynamic bounds), respectively.

Table 4. Values of $|S_k|$, $w_k$, $c_k$, $W_k$, and $C_k$ for $k$ = 1, 2, …., 14.

| k | Original | | | | | R | | | | | RN | RNWC |
|---|---|---|---|---|---|---|---|---|---|---|---|---|
| | $|S_k|$ | $w_k$ | $c_k$ | $W_k$ | $C_k$ | $|S_k|$ | $w_k$ | $c_k$ | $W_k$ | $C_k$ | $|S_k|$ | $|S_k|$ |
| 1 | 494 | 2 | 1 | 66 | 33 | 490 | 4 | 2 | 109 | 57 | 490 | 193 |
| 2 | 164 | 8 | 1 | 58 | 32 | 161 | 16 | 2 | 93 | 55 | 96 | 96 |
| 3 | 494 | 4 | 1 | 54 | 31 | 490 | 8 | 2 | 85 | 53 | 488 | 105 |
| 4 | 164 | 4 | 3 | 50 | 28 | 161 | 8 | 6 | 77 | 47 | 161 | 151 |
| 5 | 164 | 3 | 2 | 47 | 26 | 161 | 6 | 4 | 71 | 43 | 161 | 105 |
| 6 | 494 | 4 | 2 | 43 | 24 | 494 | 4 | 2 | 67 | 41 | 494 | 267 |
| 7 | 164 | 7 | 4 | 36 | 20 | 161 | 14 | 8 | 53 | 33 | 121 | 105 |
| 8 | 164 | 4 | 3 | 32 | 17 | 161 | 8 | 6 | 45 | 27 | 159 | 61 |
| 9 | 494 | 7 | 2 | 25 | 15 | 493 | 7 | 2 | 38 | 25 | 229 | 54 |
| 10 | 164 | 5 | 4 | 20 | 11 | 161 | 10 | 8 | 28 | 17 | 161 | 42 |
| 11 | 164 | 5 | 3 | 15 | 8 | 162 | 6 | 5 | 22 | 12 | 162 | 66 |
| 12 | 494 | 4 | 2 | 11 | 6 | 490 | 8 | 4 | 14 | 8 | 488 | 84 |
| 13 | 164 | 5 | 2 | 6 | 4 | 164 | 5 | 2 | 9 | 6 | 164 | 44 |
| 14 | 494 | 6 | 4 | | | 491 | 9 | 6 | | | 400 | 82 |
| product | 7.6052E+33 | | | | | 6.5097E+33 | | | | | 1.0813E+33 | 2.3930E+27 |

**4.4 Proposed Dominance Rule**

A vector $X \in \bigotimes_{i=1}^{k} S_i = S_1 \otimes S_2 \otimes \ldots \otimes S_k$ is dominated by vector $X^* \in \bigotimes_{i=1}^{k} S_i$ if $W(X^*) \leq W(X)$, $C(X^*) \leq C(X)$, and $R(X^*) \geq R(X)$. The above action of finding and removing dominated vectors is called the dominated rule. The role of the dominance rule is critical to the proposed BRB in reducing the size of $S_i$ for all $i \in \Omega$.

From Lemma 2, the size of ($S_1 \otimes S_2 \otimes \ldots \otimes S_{14}$) can be reduced if the size of $\bigotimes_{i=1}^{k} S_i$ can be removed for $k$ = 1, 2, …, 14. The following lemma proves that all dominated vectors in $\bigotimes_{i=1}^{k} S_i$ can be removed without losing any optimum.

**Lemma 5.** The vector $\bigotimes_{i=1}^{k} X_i = X_1 \otimes X_2 \otimes \ldots \otimes X_k \in \bigotimes_{i=1}^{k} S_i$ is dominated, $\bigotimes_{i=1}^{k} X_i$ can be removed from $\bigotimes_{i=1}^{k} S_i$ without losing the optimum for all $X_i \in S_i$ and $i = 1, 2, \ldots, k$.

**Proof.**

Let $X = \bigotimes_{i=1}^{k} X_i \in \bigotimes_{i=1}^{k} S_i$ be dominated by $X^* = \bigotimes_{i=1}^{k} X_i^* \in \bigotimes_{i=1}^{k} S_i$, i.e., $W(X^*) \leq W(X)$, $C(X^*) \leq C(X)$, and $R(X^*) \geq R(X)$. The solution $[\bigotimes_{i=1}^{k} X_i] \otimes [\bigotimes_{i=k+1}^{n} X_i] = X \otimes [\bigotimes_{i=k+1}^{n} X_i] \in \bigotimes_{i=1}^{n} S_i$ is either infeasible or not an optimum solution because

$$W(X^* \otimes [\bigotimes_{i=k+1}^{n} X_i]) = W(X^*) + W(\bigotimes_{i=k+1}^{n} X_i) \leq W(X \otimes [\bigotimes_{i=k+1}^{n} X_i]) = W(X) + W(\bigotimes_{i=k+1}^{n} X_i), \qquad (24)$$

$$C(X^* \otimes [\bigotimes_{i=k+1}^{n} X_i]) = C(X^*) + C(\bigotimes_{i=k+1}^{n} X_i) \leq C(X \otimes [\bigotimes_{i=k+1}^{n} X_i]) = C(X) + C(\bigotimes_{i=k+1}^{n} X_i), \qquad (25)$$

$$R(X^* \otimes [\bigotimes_{i=k+1}^{n} X_i]) = R(X^*) \times R(\bigotimes_{i=k+1}^{n} X_i) \geq R(X \otimes [\bigotimes_{i=k+1}^{n} X_i]) = R(X) \times R(\bigotimes_{i=k+1}^{n} X_i). \qquad (26)$$

Hence, this lemma is correct. □

Let $D(\bullet)$ be the new $\bullet$ after using the proposed dominated rule, that is, find and remove all dominated vectors. We have the following corollary.

**Corollary 5.** $\bigotimes_{i=1}^{k} D(S_i) = D(\bigotimes_{i=1}^{k} S_i)$, where $D(S_i)$ is the new $S_i$ after implementing the dominance rule.

### 4.5 Dynamic Bounds: $W_i$ and $C_i$

Based on the limitations of the weight and cost in Eqs. (10) and (11) together with the best-known solutions, there are two type of dynamic bounds ($W_i$ and $C_i$ for $i = 1, 2, \ldots, 14$) and one fixed bound ($R_{LB}$) in the proposed algorithm. The following lemma discusses how to implement the dynamic weight and cost bounds to detect infeasible vectors in the earlier stage.

**Lemma 6.** $\bigotimes_{i=1}^{k} X_i$ is infeasible if any of the following equations is dissatisfied:

$$W_{UB} \leq W(\bigotimes_{i=1}^{k} X_i) + W_k \qquad (27)$$

$$C_{UB} \leq C(\bigotimes_{i=1}^{k} X_i) + C_k \qquad (28)$$

$$R(\bigotimes_{i=1}^{k} X_i) \leq R_{UB}. \qquad (29)$$

**Proof.**

Follows from Lemma 3 and Eqs. (10)–(11) directly. □

Let $B(\bullet)$ be the new $\bullet$ after using the proposed dynamic bounds and $R_{LB}$ to find and remove all infeasible vectors. From the above discussion, we have the following theorem:

**Theorem 1.** $B(\bigotimes_{i=1}^{k} S_i) = \bigotimes_{i=1}^{k} B(S_i)$ and $\bigotimes_{i=1}^{k} D(B(S_i)) = B(\bigotimes_{i=1}^{k} D(S_i))$.

### 4.6 Overall Procedure of the proposed BRB

According to discussions in the previous subsection of Section 4, the procedure of overall proposed BRB is described as follows.

**Algorithm: the BRB**

**Input:** Fyffe RAP.

**Output:** Th exact solution of the Fyffe RAP.

**STEP 0.** Find all vectors in $B_4$ using the proposed upper-bound BAT.

**STEP 1.** Find $B_3$ from $B_4$ based on Lemma 1.

**STEP 2.** Let $S_i = B_3$ and $B_4$ for $i \in \Omega_3$ and $\Omega_4$, respectively.

**STEP 3.** Remove these vectors with reliabilities less than or equal to $R_{LB}$ from $S_i$ for $i = 1, 2, \ldots, 14$.

**STEP 4.** Remove all dominated vectors from $S_i$ for $i = 1, 2, \ldots, 14$.

**STEP 5.** Let $i = 2$, $j = k = 1$, $S_{new} = \emptyset$, and $S = S_1$.

**STEP 6.** Let $X = X_j \otimes X_{i,k}$, where $X_j$ is the $j$th vector in $S$.

**STEP 7.** If $W_{UB} < W(X) + W_i$, $C_{UB} < C(X) + C_i$, or $R(X) < R_{LB}$, discard $X$. Otherwise, let $S_{new} = S_{new} \cup \{X\}$.

**STEP 8.** If $j < |S|$, let $j = j + 1$, and go to STEP 6.

**STEP 9.** If $k < |S_i|$, let $j = 1$ and $k = k + 1$ and go to STEP 6.

**STEP 10.** $S = \{X \mid \text{for all } X \text{ is nondominated in } S_{new}\}$.

**STEP 11.** If $i < 14$, let $i = i + 1$, $j = k = 1$, $S_{new} = \emptyset$, and go to STEP 6. Otherwise, the best vector with largest reliability in $S$ is the optimum.

The BRB implements the upper-bound BAT to find $B_4$ in STEP 0. The concept of the super sub-BAT is implemented in STEP 1 to have $B_4$ in STEP 1 and to split the Fyffe RAP into smaller-size sub-problems by letting $S_i = B_3$ and $B_4$ for $i \in \Omega_3$ and $\Omega_4$, respectively. The multiplication of super sub-BATs is used to have the convolutional subsystem $\bigotimes_{i=1}^{k} S_i$ for $k = 1, 2, \ldots, 14$ in STEP 6. The dynamic bounds (in STEP 7), the proposed dominance rule (in STEPs 4 and 10), and $R_{LB}$ (in STEPs 3 and 7) are employed to reduce the size of the convolutional subsystem $\bigotimes_{i=1}^{k} S_i$ for $k = 1, 2, \ldots, 14$.

## 5. SOLVE FYFFE RAP

The proposed BRB is applied to the 33-variation Fyffe RAP which is the most famous benchmark RAP. The proposed BRB were programmed in C$^{++}$ language, run on an Intel(R) Core(TM) i7-10750H CPU @ 2.60GHz & 2.59 GHz with 64 GB memory, conducted on Windows 11 enterprise, and the runtime is recorded in CPU seconds.

Table 5. Optimal solutions obtained from BRB.

| ID | 1 | 2 | 3 | 4 | 5 | 6 | 7 | 8 | 9 | 10 | 11 | 12 | 13 | 14 |
|---|---|---|---|---|---|---|---|---|---|---|---|---|---|---|
| 1  | 0.95456482340 | 0030 | 200 | 0002 | 003 | 020 | 0200 | 200 | 300 | 0020 | 030 | 200 | 4000 | 020 | 0020 |
| 2  | 0.95571443870 | 0030 | 200 | 0002 | 003 | 020 | 0200 | 200 | 300 | 0020 | 030 | 101 | 4000 | 020 | 0020 |
| 3  | 0.95803460230 | 0030 | 200 | 0002 | 003 | 020 | 0200 | 200 | 201 | 0020 | 030 | 200 | 4000 | 020 | 0020 |
| 4  | 0.95918839640 | 0030 | 200 | 0002 | 003 | 020 | 0200 | 200 | 201 | 0020 | 030 | 101 | 4000 | 020 | 0020 |
| 5  | 0.96064241580 | 0030 | 200 | 0002 | 003 | 020 | 0200 | 101 | 201 | 0020 | 030 | 200 | 4000 | 020 | 0020 |
| 6  | 0.96242186270 | 0030 | 200 | 0002 | 003 | 030 | 0200 | 200 | 201 | 0020 | 030 | 200 | 4000 | 020 | 0020 |
| 7  | 0.96371184210 | 0030 | 200 | 0003 | 003 | 020 | 0200 | 200 | 201 | 0020 | 030 | 200 | 4000 | 020 | 0020 |
| 8  | 0.96504161850 | 0030 | 200 | 0002 | 003 | 030 | 0200 | 101 | 201 | 0020 | 030 | 200 | 4000 | 020 | 0020 |
| 9  | 0.96633510930 | 0030 | 200 | 0003 | 003 | 020 | 0200 | 101 | 201 | 0020 | 030 | 200 | 4000 | 020 | 0020 |
| 10 | 0.96812510110 | 0030 | 200 | 0003 | 003 | 030 | 0200 | 200 | 201 | 0020 | 030 | 200 | 4000 | 020 | 0020 |
| 11 | 0.96929104760 | 0030 | 200 | 0003 | 003 | 030 | 0200 | 200 | 201 | 0020 | 030 | 101 | 4000 | 020 | 0020 |
| 12 | 0.97076038140 | 0030 | 200 | 0003 | 003 | 030 | 0200 | 101 | 201 | 0020 | 030 | 200 | 4000 | 020 | 0020 |
| 13 | 0.97192950160 | 0030 | 200 | 0003 | 003 | 030 | 0200 | 101 | 201 | 0020 | 030 | 101 | 4000 | 020 | 0020 |
| 14 | 0.97302662370 | 0030 | 200 | 0003 | 003 | 030 | 0200 | 101 | 201 | 0020 | 021 | 101 | 4000 | 020 | 0020 |
| 15 | 0.97382683580 | 0030 | 200 | 0003 | 003 | 030 | 0200 | 101 | 400 | 0020 | 030 | 101 | 4000 | 020 | 0020 |
| 16 | 0.97492609960 | 0030 | 200 | 0003 | 003 | 030 | 0200 | 101 | 400 | 0020 | 021 | 101 | 4000 | 020 | 0020 |
| 17 | 0.97570791650 | 0030 | 200 | 0003 | 003 | 030 | 0200 | 101 | 400 | 0020 | 021 | 002 | 4000 | 020 | 0020 |
| 18 | 0.97669049270 | 0030 | 200 | 0003 | 003 | 030 | 0200 | 002 | 400 | 0020 | 021 | 101 | 4000 | 020 | 0020 |
| 19 | 0.97759630660 | 0030 | 200 | 0003 | 003 | 030 | 0200 | 300 | 201 | 0020 | 021 | 101 | 4000 | 020 | 0020 |
| 20 | 0.97840027680 | 0030 | 200 | 0003 | 003 | 030 | 0200 | 300 | 400 | 0020 | 030 | 101 | 4000 | 020 | 0020 |
| 21 | 0.97950470320 | 0030 | 200 | 0003 | 003 | 030 | 0200 | 300 | 400 | 0020 | 021 | 101 | 4000 | 020 | 0020 |
| 22 | 0.98029019180 | 0030 | 200 | 0003 | 003 | 030 | 0200 | 300 | 400 | 0020 | 021 | 002 | 4000 | 020 | 0020 |
| 23 | 0.98102706670 | 0030 | 200 | 0003 | 003 | 030 | 0200 | 300 | 400 | 0020 | 012 | 002 | 4000 | 020 | 0020 |
| 24 | 0.98151831690 | 0030 | 200 | 0003 | 003 | 030 | 0200 | 300 | 400 | 0020 | 003 | 002 | 4000 | 020 | 0020 |
| 25 | 0.98225568740 | 0030 | 200 | 0003 | 003 | 030 | 0200 | 300 | 400 | 0020 | 021 | 002 | 4000 | 020 | 0011 |
| 26 | 0.98299403980 | 0030 | 200 | 0003 | 003 | 030 | 0200 | 300 | 400 | 0020 | 012 | 002 | 4000 | 020 | 0011 |
| 27 | 0.98350485170 | 0030 | 200 | 0003 | 004 | 030 | 0200 | 300 | 400 | 0110 | 021 | 101 | 4000 | 020 | 0020 |
| 28 | 0.98417552490 | 0030 | 200 | 0003 | 003 | 030 | 0200 | 300 | 400 | 0110 | 012 | 002 | 4000 | 020 | 0011 |
| 29 | 0.98468809620 | 0030 | 200 | 0003 | 004 | 030 | 0200 | 300 | 400 | 1010 | 021 | 101 | 4000 | 020 | 0011 |
| 30 | 0.98537823540 | 0030 | 200 | 0003 | 004 | 030 | 0200 | 300 | 400 | 0110 | 021 | 101 | 4000 | 110 | 0011 |
| 31 | 0.98592167210 | 0030 | 200 | 0003 | 004 | 030 | 0200 | 300 | 400 | 0110 | 012 | 101 | 4000 | 200 | 0011 |
| 32 | 0.98641607690 | 0030 | 200 | 0003 | 004 | 030 | 0200 | 300 | 400 | 2000 | 012 | 002 | 4000 | 110 | 0011 |
| 33 | 0.98681101780 | 0030 | 200 | 0003 | 004 | 030 | 0200 | 300 | 400 | 1100 | 012 | 002 | 4000 | 200 | 0011 |

Without using the proposed dominance rule or two types of dynamic bounds, the computer memory is crashed due to the overflow number of solutions. Hence, the dominance rule or two dynamic bounds are must-have. In the experiment, we also test the role of $R_{LB}$ and the value of $R_{UB}$ for each variant in the Fyffe RAP adapted from the results obtained from the

SSO [20].

Table 5 provides the exact solutions obtained from the proposed BRB. Tables 6 and 7 show the value of $\Delta i = |\{X \in \bigotimes_{k=1}^{j} S_k \text{ for ID}=i\}| - |\{X \in \bigotimes_{k=1}^{j} S_k \text{ for ID}=(i+1)\}|$ for $i = 1, 2, \ldots, 32$ and $j = 1, 2, \ldots, 14$ after using the dominance rule, two dynamic bounds, and $R_{LB}$ (used in Table 6 only). Note that $W_{UB} = 159 + (i-1)$ and $C_{UB} = 130$ for ID $= i$ and $i = 1, 2, \ldots, 33$.

Figures 2 and 3 depict the values of $|\bigotimes_{k=1}^{i} S_k|$ to check the role of $R_{LB}$ before and after using the dominance rule for $i = 1, 2, \ldots, 13$, respectively. Figures 4-7 compare the results between the combinations of $R_{LB}$ and the dominance rule with only $R_{LB}$ or only the dominance rule.

**Table 6.** Results from the proposed algorithm under $R_{LB}$[&]

| i | 0 | 1 | Δ2 | Δ3 | Δ4 | Δ5 | Δ6 | Δ7 | Δ8 | Δ9 | Δ10 | Δ11 | Δ12 | Δ13 | Δ14 | Δ15 | Δ16 | Δ17 | Δ18 | Δ19 | Δ20 | Δ21 | Δ22 | Δ23 | Δ24 | Δ25 | Δ26 | Δ27 | Δ28 | Δ29 | Δ30 | Δ31 | Δ32 | Δ33 |
|---|---|---|---|---|---|---|---|---|---|---|---|---|---|---|---|---|---|---|---|---|---|---|---|---|---|---|---|---|---|---|---|---|---|---|
| 1* | 490 | 490 | 0 | 0 | 0 | 0 | 0 | 0 | 0 | 0 | 0 | 0 | 0 | 0 | 0 | 0 | 0 | 0 | 0 | 0 | 0 | 0 | 0 | 0 | 0 | 0 | 0 | 0 | 0 | 0 | 0 | 0 | 0 | 0 |
| 1# | 490 | 490 | 0 | 0 | 0 | 0 | 0 | 0 | 0 | 0 | 0 | 0 | 0 | 0 | 0 | 0 | 0 | 0 | 0 | 0 | 0 | 0 | 0 | 0 | 0 | 0 | 0 | 0 | 0 | 0 | 0 | 0 | 0 | 0 |
| 2* | 161 | 9070 | 486 | 509 | 519 | 545 | 555 | 582 | 592 | 616 | 625 | 650 | 658 | 682 | 690 | 711 | 715 | 731 | 728 | 737 | 727 | 726 | 710 | 701 | 677 | 660 | 630 | 604 | 567 | 535 | 493 | 457 | 416 | 378 |
| 2# | 161 | 483 | 16 | 17 | 15 | 16 | 14 | 15 | 15 | 16 | 15 | 15 | 15 | 15 | 15 | 15 | 15 | 15 | 14 | 15 | 12 | 13 | 13 | 12 | 12 | 12 | 13 | 13 | 11 | 13 | 13 | 14 | 14 | 14 |
| 3* | 490 | 21614 | 1153 | 1220 | 1206 | 1268 | 1248 | 1312 | 1301 | 1369 | 1356 | 1413 | 1392 | 1441 | 1421 | 1475 | 1456 | 1507 | 1483 | 1529 | 1506 | 1543 | 1516 | 1533 | 1536 | 1553 | 1440 | 927 | 1513 | 1515 | 1422 | 1529 | 1492 | 1386 |
| 3# | 490 | 949 | 27 | 30 | 28 | 34 | 31 | 36 | 36 | 37 | 37 | 39 | 38 | 35 | 33 | 35 | 32 | 36 | 35 | 37 | 32 | 33 | 34 | 39 | 37 | 38 | 37 | 40 | 38 | 40 | 38 | 42 | 39 | 39 |
| 4* | 161 | 57036 | 3220 | 3291 | 3341 | 3444 | 3432 | 3585 | 3589 | 3697 | 3600 | 3634 | 3486 | 2698 | 3629 | 3705 | 2367 | 3987 | 3793 | 2456 | 2855 | 3467 | 3735 | 3095 | 4086 | 3994 | 3868 | 3429 | 4349 | 4457 | 4268 | 4663 | 4347 | 4398 |
| 4# | 161 | 1201 | 33 | 28 | 34 | 35 | 35 | 39 | 39 | 40 | 33 | 38 | 39 | 42 | 43 | 43 | 39 | 43 | 43 | 41 | 44 | 44 | 46 | 47 | 45 | 49 | 49 | 52 | 50 | 49 | 45 | 40 | 45 | 39 |
| 5* | 161 | 59558 | 2714 | 1992 | 2578 | 2433 | 2483 | 2837 | 2775 | 2870 | 2309 | 3008 | 3038 | 3155 | 3399 | 3507 | 3545 | 3826 | 3675 | 3646 | 3995 | 3840 | 4003 | 3961 | 3791 | 4018 | 4123 | 4390 | 4283 | 4268 | 3849 | 3814 | 4226 | 3689 |
| 5# | 161 | 1361 | 45 | 25 | 39 | 33 | 33 | 32 | 33 | 34 | 40 | 41 | 41 | 45 | 45 | 46 | 39 | 43 | 40 | 44 | 47 | 46 | 48 | 42 | 44 | 37 | 39 | 47 | 46 | 51 | 39 | 42 | 39 | 37 |
| 6* | 494 | **128209** | **7065** | **4427** | **7227** | **6134** | **6342** | **6265** | **6845** | **6770** | **7480** | **7250** | **7666** | **8715** | **8837** | **9250** | **7764** | **8909** | **7968** | **8533** | **9385** | **8859** | **8898** | **8548** | **8991** | **7790** | **7906** | **10326** | **9869** | **10962** | **7822** | **8588** | **8124** | **7118** |
| 6# | 494 | 1270 | 34 | 34 | 42 | 39 | 40 | 41 | 41 | 43 | 39 | 43 | 39 | 36 | 37 | 39 | 38 | 42 | 41 | 47 | 46 | 37 | 37 | 39 | 42 | 41 | 46 | 42 | 38 | 39 | 28 | 35 | 41 | 33 |
| 7* | 161 | 32010 | 1682 | 1488 | 2094 | 2113 | 2028 | 2263 | 1993 | 2174 | 1837 | 2825 | 2465 | 2608 | 2455 | 2756 | 2205 | 2798 | 3062 | 3570 | 3679 | 2862 | 2764 | 2797 | 3375 | 3125 | 3718 | 3760 | 3416 | 3723 | 2427 | 2971 | 3845 | 3126 |
| 7# | 161 | 1195 | 37 | 21 | 33 | 37 | 36 | 35 | 31 | 37 | 24 | 42 | 39 | 35 | 30 | 38 | 32 | 37 | 33 | 31 | 24 | 28 | 31 | 35 | 30 | 29 | 26 | 28 | 23 | 32 | 15 | 32 | 32 | 17 |
| 8* | 161 | 14681 | 890 | 366 | 672 | 769 | 749 | 875 | 814 | 781 | 239 | 1153 | 1077 | 869 | 681 | 954 | 646 | 944 | 726 | 696 | 491 | 232 | 332 | 653 | 826 | 788 | 492 | 1095 | 742 | 919 | 549 | 838 | 963 | 747 |
| 8# | 161 | 1023 | 34 | 17 | 31 | 21 | 25 | 35 | 28 | 19 | 13 | 27 | 6 | 13 | 14 | 27 | 5 | 24 | 10 | 20 | 8 | 17 | 18 | 27 | 36 | 17 | 26 | 27 | 20 | 31 | 16 | 18 | 29 | 24 |
| 9* | 493 | 10397 | 715 | 324 | 622 | 331 | 445 | 762 | 582 | 320 | 41 | 522 | -20 | 248 | 208 | 554 | 169 | 511 | 242 | 370 | 407 | 122 | 482 | 521 | 874 | 597 | 483 | 657 | 579 | 778 | 403 | 613 | 823 | 699 |
| 9# | 493 | 844 | 22 | -1 | 24 | 9 | 9 | 21 | 14 | 8 | 3 | 14 | -5 | 6 | 8 | 17 | -1 | 15 | 6 | -1 | 19 | -5 | 16 | 16 | 30 | 15 | 5 | 26 | 17 | 17 | 14 | 22 | 23 | 24 |
| 10* | 161 | 7873 | 395 | -237 | 319 | 90 | -7 | 298 | 292 | 102 | -175 | 153 | -156 | 28 | 97 | 272 | 15 | 367 | 131 | 92 | 288 | -123 | 195 | 293 | 566 | 238 | 67 | 427 | 235 | 389 | 65 | 310 | 437 | 276 |
| 10# | 161 | 611 | 12 | -12 | 13 | 15 | 3 | 11 | 15 | 3 | -15 | 11 | -2 | 4 | 8 | 14 | 4 | 14 | 4 | 7 | 13 | 0 | 6 | 10 | 20 | 13 | 1 | 14 | 13 | 14 | 3 | 15 | 23 | 16 |
| 11* | 162 | 4710 | 210 | -198 | 232 | 269 | 26 | 56 | 166 | 70 | -242 | 112 | -86 | -36 | 151 | 293 | 54 | 274 | 88 | 138 | 298 | 20 | 169 | 263 | 434 | 288 | 65 | 357 | 278 | 355 | 128 | 300 | 522 | 238 |
| 11# | 162 | 403 | 14 | -2 | 14 | 5 | -3 | 1 | 7 | 0 | -12 | 11 | -11 | 7 | 7 | 11 | 4 | 10 | 3 | 3 | 7 | -8 | 3 | 9 | 16 | 1 | -1 | 12 | -1 | 14 | -4 | 5 | 12 | 8 |
| 12* | 490 | 1867 | 149 | -112 | 89 | 6 | -84 | 29 | 16 | 30 | -181 | 82 | -80 | 1 | -8 | 115 | 51 | 136 | 37 | 34 | 67 | -176 | 7 | 35 | 202 | 51 | -17 | 146 | 18 | 115 | -108 | -24 | 137 | 4 |
| 12# | 490 | 137 | 12 | -8 | 7 | 2 | -7 | 1 | -2 | 4 | -4 | 3 | -8 | -4 | 6 | 9 | 2 | 8 | 0 | 1 | 4 | -4 | 6 | 3 | 10 | -3 | -2 | 11 | -6 | 8 | -6 | 3 | 6 | 5 |
| 13* | 164 | 140 | 30 | -29 | 21 | 2 | -24 | 11 | 3 | -1 | -10 | 0 | -20 | -6 | 22 | 24 | 7 | 12 | 0 | 2 | -4 | -6 | 1 | 20 | 38 | 18 | 3 | 14 | -6 | 27 | -7 | 1 | 11 | -8 |
| 13# | 164 | 47 | 10 | -9 | 7 | 1 | -6 | 4 | 3 | -1 | -2 | 5 | -5 | -1 | 7 | 5 | 3 | 1 | 0 | 2 | 0 | -1 | 2 | 8 | 9 | 4 | -2 | 5 | -1 | 5 | -7 | -3 | 3 | 1 |
| 14* | 491 | 1 | 0 | 0 | 0 | 0 | 0 | 0 | 0 | 0 | 0 | 0 | 0 | 0 | 0 | 0 | 0 | 0 | 0 | 0 | 0 | 0 | 0 | 0 | 1 | -1 | 0 | 0 | 0 | 0 | 0 | 0 | 0 | 0 |
| 14# | 491 | 1 | 0 | 0 | 0 | 0 | 0 | 0 | 0 | 0 | 0 | 0 | 0 | 0 | 0 | 0 | 0 | 0 | 0 | 0 | 0 | 0 | 0 | 0 | 0 | 0 | 0 | 0 | 0 | 0 | 0 | 0 | 0 | 0 |

[&] The values in bold and square are the largest among related column and row, respectively.
[*] before using the dominance rule
[#] after using the dominance rule

**Table 7.** Results from the proposed algorithm without using $R_{LB}$[&]

| i | 0 | 1 | Δ2 | Δ3 | Δ4 | Δ5 | Δ6 | Δ7 | Δ8 | Δ9 | Δ10 | Δ11 | Δ12 | Δ13 | Δ14 | Δ15 | Δ16 | Δ17 | Δ18 | Δ19 | Δ20 | Δ21 | Δ22 | Δ23 | Δ24 | Δ25 | Δ26 | Δ27 | Δ28 | Δ29 | Δ30 | Δ31 | Δ32 | Δ33 |
|---|---|---|---|---|---|---|---|---|---|---|---|---|---|---|---|---|---|---|---|---|---|---|---|---|---|---|---|---|---|---|---|---|---|---|
| 1[*] | 490 | 490 | 0 | 0 | 0 | 0 | 0 | 0 | 0 | 0 | 0 | 0 | 0 | 0 | 0 | 0 | 0 | 0 | 0 | 0 | 0 | 0 | 0 | 0 | 0 | 0 | 0 | 0 | 0 | 0 | 0 | 0 | 0 | 0 |
| 1[#] | 490 | 490 | 0 | 0 | 0 | 0 | 0 | 0 | 0 | 0 | 0 | 0 | 0 | 0 | 0 | 0 | 0 | 0 | 0 | 0 | 0 | 0 | 0 | 0 | 0 | 0 | 0 | 0 | 0 | 0 | 0 | 0 | 0 | 0 |
| 2[*] | 161 | 30841 | 240 | 210 | 183 | 157 | 135 | 114 | 96 | 79 | 65 | 52 | 41 | 31 | 23 | 16 | 11 | 7 | 4 | 2 | 1 | 0 | 0 | 0 | 0 | 0 | 0 | 0 | 0 | 0 | 0 | 0 | 0 | 0 |
| 2[#] | 161 | 1022 | 15 | 15 | 16 | 15 | 14 | 13 | 12 | 11 | 10 | 9 | 9 | 6 | 4 | 4 | 2 | 1 | 1 | 0 | 0 | 0 | 0 | 0 | 0 | 0 | 0 | 0 | 0 | 0 | 0 | 0 | 0 | 0 |
| 3[*] | 490 | 73874 | 1591 | 1567 | 1583 | 1554 | 1564 | 1538 | 1550 | 1524 | 1530 | 1501 | 1505 | 1474 | 1473 | 1443 | 1442 | 1410 | 1402 | 1369 | 1362 | 1329 | 1318 | 1282 | 1270 | 1234 | 1219 | 1180 | 1161 | 1120 | 1096 | 1053 | 1026 | 981 |
| 3[#] | 490 | 2252 | 44 | 41 | 43 | 40 | 44 | 42 | 42 | 41 | 40 | 39 | 41 | 37 | 39 | 39 | 38 | 39 | 39 | 38 | 40 | 36 | 39 | 38 | 39 | 37 | 40 | 36 | 40 | 38 | 41 | 41 | 43 | 41 |
| 4[*] | 161 | 197064 | 5207 | 5223 | 5331 | 5351 | 5455 | 5477 | 5579 | 5600 | 5693 | 5707 | 5803 | 5814 | 5897 | 5901 | 5979 | 5979 | 6050 | 6042 | 6106 | 6093 | 6150 | 6142 | 6194 | 6180 | 6231 | 6213 | 6263 | 6240 | 6284 | 6259 | 6294 | 6256 |
| 4[#] | 161 | 2792 | 55 | 55 | 54 | 54 | 54 | 56 | 59 | 58 | 60 | 57 | 59 | 58 | 59 | 58 | 59 | 59 | 60 | 58 | 57 | 56 | 56 | 57 | 58 | 55 | 57 | 55 | 56 | 54 | 54 | 52 | 52 | 51 |
| 5[*] | 161 | 194396 | 4776 | 4834 | 4871 | 4917 | 4962 | 5020 | 5077 | 5132 | 5178 | 5239 | 5290 | 5339 | 5381 | 5433 | 5480 | 5525 | 5564 | 5610 | 5651 | 5697 | 5736 | 5782 | 5819 | 5859 | 5893 | 5925 | 5951 | 5970 | 5980 | 5985 | 5983 | 5973 |
| 5[#] | 161 | 3010 | 58 | 57 | 54 | 55 | 57 | 57 | 57 | 60 | 61 | 61 | 60 | 61 | 62 | 61 | 64 | 61 | 62 | 61 | 62 | 66 | 65 | 65 | 65 | 65 | 66 | 65 | 64 | 64 | 64 | 64 | 63 | 63 |
| 6[*] | **494** | **504109** | 12072 | 12206 | 12313 | 12408 | 12537 | 12626 | 12716 | 12782 | 12890 | 13022 | 13066 | 13140 | 13241 | 13339 | 13485 | 13597 | 13753 | 13917 | 14075 | 14222 | 14367 | 14474 | 14573 | 14679 | 14762 | 14827 | 14881 | 14960 | 15006 | 15029 | 15023 | 15013 |
| 6[#] | 494 | 3095 | 55 | 54 | 55 | 55 | 57 | 59 | 60 | 59 | 58 | 59 | 63 | 62 | 57 | 57 | 61 | 61 | 61 | 61 | 65 | 64 | 64 | 67 | 67 | 66 | 67 | 68 | 65 | 64 | 62 | 64 | 63 | 62 |
| 7[*] | 161 | 183528 | 5360 | 5425 | 5496 | 5555 | 5622 | 5693 | 5761 | 5834 | 5891 | 5952 | 6011 | 6081 | 6136 | 6201 | 6273 | 6333 | 6398 | 6444 | 6506 | 6555 | 6610 | 6682 | 6749 | 6838 | 6910 | 6992 | 7047 | 7106 | 7160 | 7219 | 7285 | 7340 |
| 7[#] | 161 | 2904 | 53 | 52 | 55 | 58 | 55 | 55 | 49 | 53 | 53 | 48 | 53 | 52 | 53 | 52 | 54 | 55 | 56 | 56 | 55 | 56 | 56 | 57 | 58 | 58 | 58 | 57 | 57 | 58 | 59 | 58 | 59 | 59 |
| 8[*] | 161 | 99522 | 2482 | 2511 | 2548 | 2573 | 2593 | 2630 | 2652 | 2675 | 2692 | 2715 | 2745 | 2761 | 2768 | 2786 | 2817 | 2826 | 2837 | 2861 | 2889 | 2901 | 2917 | 2933 | 2963 | 2978 | 2984 | 2998 | 3014 | 3020 | 3013 | 3005 | 3004 | 2986 |
| 8[#] | 161 | 2785 | 48 | 47 | 49 | 50 | 51 | 49 | 53 | 51 | 51 | 51 | 50 | 53 | 53 | 52 | 54 | 57 | 52 | 53 | 54 | 59 | 57 | 54 | 56 | 59 | 58 | 53 | 52 | 50 | 52 | 51 | 51 | 52 |
| 9[*] | 493 | 112163 | 2835 | 2856 | 2866 | 2900 | 2933 | 2973 | 3014 | 3051 | 3093 | 3109 | 3133 | 3162 | 3206 | 3249 | 3290 | 3340 | 3370 | 3396 | 3406 | 3443 | 3479 | 3519 | 3555 | 3595 | 3634 | 3639 | 3654 | 3656 | 3683 | 3710 | 3733 | 3774 |
| 9[#] | 493 | 2778 | 46 | 44 | 45 | 45 | 45 | 46 | 47 | 48 | 48 | 48 | 48 | 49 | 50 | 51 | 50 | 53 | 51 | 53 | 51 | 50 | 54 | 52 | 52 | 54 | 54 | 56 | 55 | 54 | 56 | 60 | 56 | 55 |
| 10[*] | 161 | 77416 | 1693 | 1710 | 1724 | 1737 | 1746 | 1754 | 1767 | 1781 | 1795 | 1802 | 1807 | 1812 | 1823 | 1840 | 1848 | 1858 | 1865 | 1875 | 1892 | 1902 | 1916 | 1930 | 1939 | 1945 | 1947 | 1963 | 1971 | 1977 | 1982 | 1989 | 2003 | 2001 |
| 10[#] | 161 | 2796 | 44 | 46 | 48 | 47 | 48 | 46 | 44 | 44 | 46 | 48 | 48 | 48 | 49 | 49 | 48 | 50 | 48 | 48 | 49 | 49 | 47 | 49 | 50 | 52 | 50 | 53 | 51 | 52 | 51 | 51 | 54 | 52 |
| 11[*] | 162 | 121942 | 2550 | 2582 | 2612 | 2638 | 2663 | 2694 | 2713 | 2740 | 2764 | 2784 | 2803 | 2827 | 2847 | 2871 | 2900 | 2922 | 2934 | 2945 | 2951 | 2964 | 2972 | 2980 | 2987 | 2983 | 2992 | 2996 | 3000 | 2988 | 2991 | 2988 | 2986 | 2981 |
| 11[#] | 162 | 2792 | 45 | 46 | 46 | 47 | 46 | 46 | 44 | 45 | 45 | 45 | 47 | 49 | 50 | 50 | 51 | 48 | 46 | 46 | 48 | 49 | 51 | 50 | 51 | 51 | 50 | 49 | 49 | 49 | 50 | 49 | 51 | 50 |
| 12[*] | 490 | 154594 | 3195 | 3222 | 3244 | 3278 | 3304 | 3335 | 3364 | 3395 | 3427 | 3465 | 3498 | 3530 | 3566 | 3615 | 3645 | 3674 | 3713 | 3749 | 3778 | 3815 | 3848 | 3873 | 3910 | 3930 | 3959 | 3993 | 4002 | 3994 | 3994 | 3990 | 3977 | 3959 |
| 12[#] | 490 | 2593 | 43 | 42 | 43 | 44 | 46 | 45 | 46 | 43 | 42 | 44 | 45 | 45 | 46 | 46 | 46 | 47 | 46 | 47 | 46 | 46 | 46 | 47 | 49 | 47 | 47 | 47 | 49 | 50 | 51 | 51 | 51 | 48 |
| 13[*] | 164 | 78013 | 1418 | 1455 | 1485 | 1489 | 1514 | 1524 | 1549 | 1579 | 1584 | 1629 | 1619 | 1649 | 1694 | 1711 | 1735 | 1714 | 1747 | 1779 | 1777 | 1827 | 1813 | 1832 | 1852 | 1876 | 1913 | 1876 | 1914 | 1918 | 1928 | 1970 | 1938 | 1950 |
| 13[#] | 164 | 2666 | 41 | 44 | 44 | 47 | 47 | 46 | 47 | 46 | 47 | 48 | 47 | 49 | 46 | 46 | 48 | 49 | 47 | 50 | 51 | 51 | 50 | 49 | 50 | 49 | 49 | 49 | 50 | 52 | 49 | 50 | 50 | 50 |
| 14[*] | 491 | 159723 | 3465 | 3481 | 3542 | 3563 | 3570 | 3618 | 3609 | 3611 | 3681 | 3687 | 3702 | 3769 | 3797 | 3802 | 3864 | 3873 | 3866 | 3944 | 3971 | 3980 | 4057 | 4098 | 4095 | 4151 | 4163 | 4141 | 4177 | 4194 | 4171 | 4183 | 4185 | 4144 |
| 14[#] | 491 | 2711 | 44 | 44 | 46 | 43 | 45 | 46 | 46 | 49 | 47 | 49 | 48 | 50 | 50 | 49 | 50 | 50 | 48 | 50 | 49 | 49 | 52 | 51 | 49 | 50 | 51 | 50 | 49 | 52 | 50 | 50 | 50 | 49 |

[&] The values in bold and square are the largest among related column and row, respectively.

[*] before using the dominance rule

[#] after using the dominance rule

## 5.1 General observations

Table 6 lists the computational results from the BRB and the best-known AI method in the Fyffe RAP, that is, SSO, including the final reliability $R$, weight $W$, cost $C$, and $R_{LB}$. In Table 6, the column $T_R$ and $T$ under BRB are the runtimes (in seconds) of the BRB with and without using $R_{LB}$, respectively. The runtime $T$ for the SSO is executed SSO 60 runs to find the best solution as mentioned in [20].

**Table 6.** Results obtained from the BRB and the SSO.

| ID | $W$ | $C$ | $R$ | BD-BAT | | | | SSO |
|---|---|---|---|---|---|---|---|---|
| | | | | $T_R$ | $T$ | $T/T_R$ | $R_{LB}$ | $T$ |
| 1  | 159 | 110 | 0.95456482340 | 1.283 | 16.532 | 12.8854 | 0.954564 | 16.5  |
| 2  | 160 | 112 | 0.95571443870 | 1.390 | 17.142 | 12.3324 | 0.955713 | 12.48 |
| 3  | 161 | 113 | 0.95803460230 | 1.472 | 17.823 | 12.1080 | 0.958034 | 16.56 |
| 4  | 162 | 115 | 0.95918839640 | 1.597 | 18.536 | 11.6068 | 0.959187 | 8.82  |
| 5  | 163 | 114 | 0.96064241580 | 1.625 | 19.393 | 11.9342 | 0.960641 | 8.82  |
| 6  | 164 | 115 | 0.96242186270 | 1.722 | 20.106 | 11.6760 | 0.96242  | 9.9   |
| 7  | 165 | 117 | 0.96371184210 | 1.887 | 20.917 | 11.0848 | 0.96371  | 13.68 |
| 8  | 166 | 116 | 0.96504161850 | 2.027 | 21.694 | 10.7025 | 0.96504  | 9.96  |
| 9  | 167 | 118 | 0.96633510930 | 2.133 | 22.563 | 10.5781 | 0.966334 | 13.32 |
| 10 | 168 | 119 | 0.96812510110 | 2.268 | 23.391 | 10.3135 | 0.968124 | 10.98 |
| 11 | 169 | 121 | 0.96929104760 | 2.429 | 24.250 | 9.9835  | 0.96929  | 6.96  |
| 12 | 170 | 120 | 0.97076038140 | 2.566 | 25.156 | 9.8036  | 0.970759 | 8.52  |
| 13 | 171 | 122 | 0.97192950160 | 2.675 | 26.091 | 9.7536  | 0.971929 | 6.3   |
| 14 | 172 | 123 | 0.97302662370 | 2.909 | 26.949 | 9.2640  | 0.973026 | 7.5   |
| 15 | 173 | 122 | 0.97382683580 | 3.110 | 27.923 | 8.9785  | 0.973826 | 15.24 |
| 16 | 174 | 123 | 0.97492609960 | 3.210 | 28.987 | 9.0302  | 0.974925 | 8.22  |
| 17 | 175 | 125 | 0.97570791650 | 3.347 | 29.931 | 8.9426  | 0.975707 | 7.68  |
| 18 | 176 | 124 | 0.97669049270 | 3.538 | 30.949 | 8.7476  | 0.976689 | 9.96  |
| 19 | 177 | 126 | 0.97759630660 | 3.787 | 32.005 | 8.4513  | 0.977595 | 9.06  |
| 20 | 178 | 125 | 0.97840027680 | 3.994 | 33.199 | 8.3122  | 0.978399 | 17.58 |
| 21 | 179 | 126 | 0.97950470320 | 4.257 | 34.198 | 8.0334  | 0.979504 | 17.1  |
| 22 | 180 | 128 | 0.98029019180 | 4.515 | 35.439 | 7.8492  | 0.980289 | 14.82 |
| 23 | 181 | 128 | 0.98102706670 | 4.702 | 36.722 | 7.8099  | 0.981026 | 16.02 |
| 24 | 182 | 130 | 0.98151831690 | 4.979 | 37.807 | 7.5933  | 0.981517 | 15.36 |
| 25 | 183 | 129 | 0.98225568740 | 5.305 | 39.184 | 7.3862  | 0.982225 | 17.1  |
| 26 | 184 | 130 | 0.98299403980 | 5.545 | 40.462 | 7.2970  | 0.982993 | 14.34 |
| 27 | 185 | 130 | 0.98350485170 | 5.900 | 41.907 | 7.1029  | 0.983504 | 17.88 |
| 28 | 186 | 129 | 0.98417552490 | 6.166 | 43.238 | 7.0123  | 0.984175 | 11.7  |
| 29 | 187 | 130 | 0.98468809620 | 6.447 | 44.493 | 6.9013  | 0.984667 | 17.76 |
| 30 | 188 | 130 | 0.98537823540 | 6.775 | 45.812 | 6.7619  | 0.985377 | 16.5  |
| 31 | 189 | 130 | 0.98592167210 | 7.087 | 47.213 | 6.6619  | 0.985921 | 17.58 |
| 32 | 190 | 130 | 0.98641607690 | 7.445 | 48.702 | 6.5416  | 0.986315 | 18.18 |
| 33 | 191 | 130 | 0.98681101780 | 7.728 | 50.124 | 6.4860  | 0.98681  | 17.4  |

Note that each value of $R_{LB}$ is based on the related final solution obtained from SSO listed in [20] and can be replaced with solutions obtained from any other algorithms, e.g., ACO [5], GA [12], Tabu search [14], Simulated Annealing [16, 17], and ABC [19].

From Table 6, the cost $C$, weight $W$, and reliability $R$ of each final solution obtained the proposed BRB and SSO are all the same, for example, $W = 159$, $C = 110$, and $R = 0.95456482340$ for ID $= 1$ for both algorithms. The final solution of the BRB and the best final solution among 60 runs of the SSO are optimum. However, we can certify that the best final solution among 60 SSO runs is optimum only after the final solution is obtained from BRB. Note that the reliability of each final solution of SSO obtained in [20] is recalculated to have more digits to represent the final solution without having the round-off error.

Futhermore, the values of $W$, $C$, $R$, $T$, and $T_R$ increased from ID $= 1$ to ID $= 33$. The reason is that $W_{UB} = 110$ and $191$ at ID $= 1$ and $33$, respectively. The larger $W_{UB}$ greater the solution space, and greater solution space the more difficult the variant of the Fyffe RAP.

## 5.2 Number of Solutions of Convolutional Subsystems

Let the Y-axis be the number of obtained solutions in convolutional subsystems, that is, $|\bigotimes_{i=1}^{k} S_i|$ for $k = 1, 2, \ldots, 14$, and the X-axis be the convolutional subsystem $i$ for $i = 1, 2, \ldots, 14$. Figure 2 depicts the values of $|\bigotimes_{i=1}^{k} S_i|$ before using the dominance rule for $k = 1, 2, \ldots, 14$ and ID $= 1$. From Figure 2, there is a big gap between the results with and without using $R_{LB}$, respectively. The gap is bigger if the number of solutions in the convolutional subsystems is larger and vice versa.

From Figure 2, the peak of both lines is at the 5$^{th}$ subsystem, that is, $|\bigotimes_{k=1}^{i} S_k|$ is increased from $i$ = 1 to 5 and then decreased from $i$ = 5 to 14 for ID = 1. We have the same observation for other ID from Table 6 if the dominance rule is not considered. Therefore, most results in other figures focus only on ID = 5.

The results using $R_{LB}$ tend to increase and decrease slowly than that of the results without using $R_{LB}$ which are unpredictable. Hence, from the above, the implementation of $R_{LB}$ can reduce the number of sub-solutions.

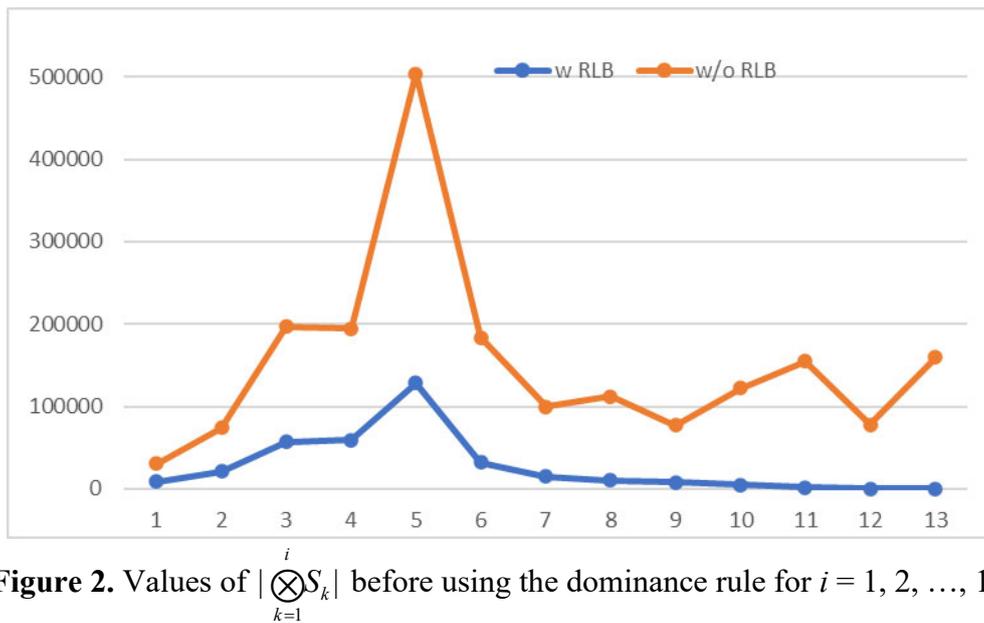

**Figure 2.** Values of $|\bigotimes_{k=1}^{i} S_k|$ before using the dominance rule for $i$ = 1, 2, …, 13.

Figure 3 shows the values of $|\bigotimes_{k=1}^{i} S_k|$ after using the dominance rule for $k$ = 1, 2, …, 14 and ID = 1. Like Figure 2, both lines are increased to their peaks at ID = 4 and 5 for results with and without $R_{LB}$, respectively, and then go down. However, the gap between two lines is grown from the first to the last subsystems most of the time and this is different to that in Figure 2. Hence, the benefit is much clearer in using $R_{LB}$ than that without using $R_{LB}$ after the dominance rule. Thus, from Figures 2 and 3, the combinations of $R_{LB}$ and the dominance rule is better than using only $R_{LB}$ or only the proposed dominance rule.

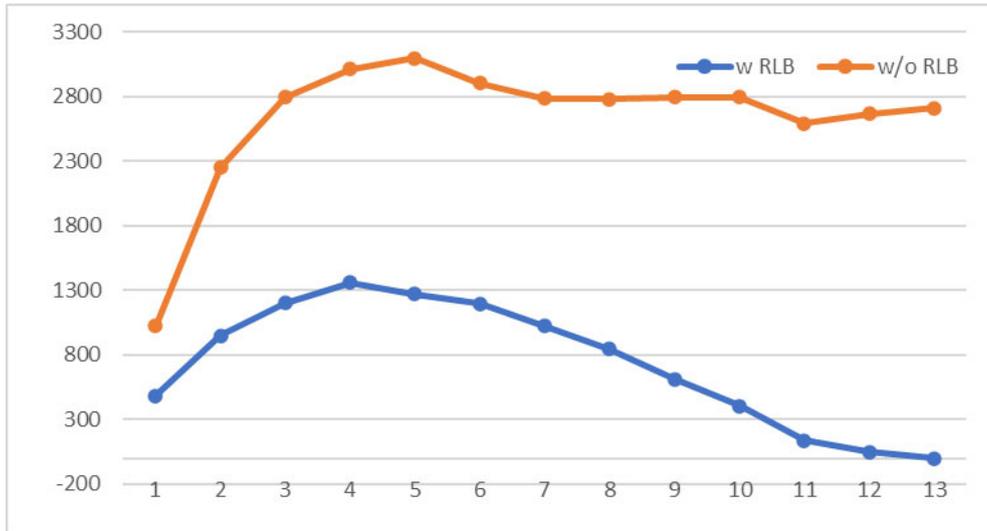

**Figure 3.** Values of $|\bigotimes_{k=1}^{i} S_k|$ after using the dominance rule for $i = 1, 2, \ldots, 13$.

### 5.3 Number of Solutions of Variants

Figures 4 and 5 show the values of $|\bigotimes_{k=1}^{5} S_k|$ after using the dominance rule for ID = 1, 2, …, 33. It can be observed that the number of solutions is increased because The larger ID greater the solution space as mentioned in subsection 5.2, the more difficult of the variant of the Fyffe RAP. However, it is interesting that both increments are linear.

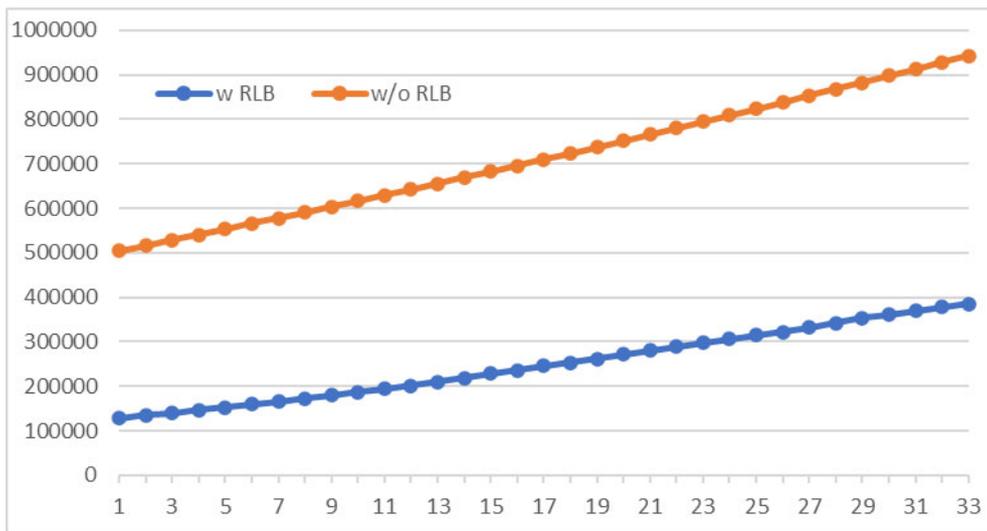

**Figure 4.** Values of $|\bigotimes_{i=1}^{5} S_k|$ before using the dominance rule for ID=1, 2, …, 33.

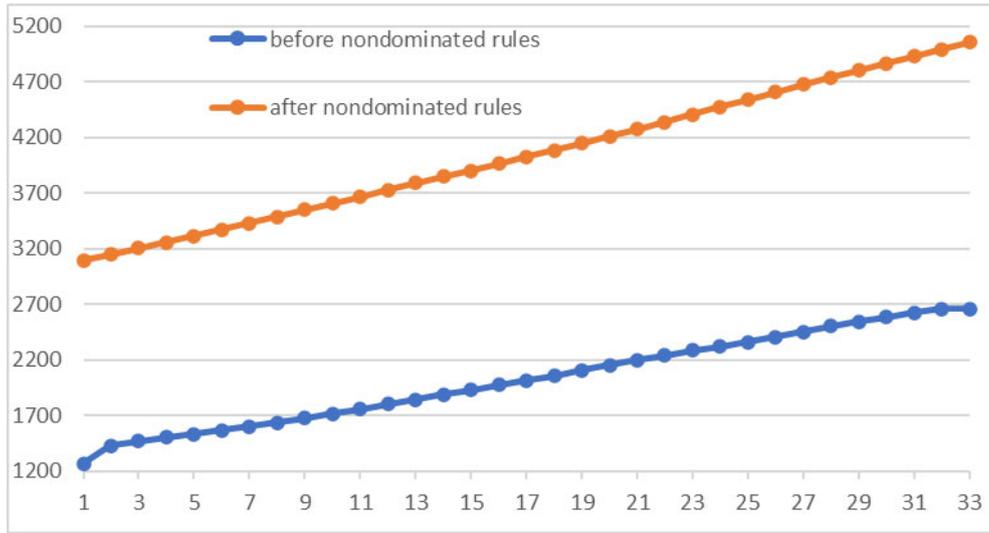

**Figure 5.** Values of $|\bigotimes_{i=1}^{5} S_k|$ after using the dominance rule for ID = 1, 2, …, 33.

Results in Figures 4 and 5 further confirm that the combinations of $R_{LB}$ and the dominance rule is better than using only $R_{LB}$ or only the dominance rule. For example, the solution number is reduced from 504,109 to 128,209 if only $R_{LB}$ is used and from 3,095 to 1,270 if $R_{LB}$ and the dominance rule are applied, respectively.

### 5.4 Ratios of Number of Solutions

Figures 6 and 7 discuss the ratios of the solution numbers obtained to further demonstrate the performance of the $R_{LB}$ and the dominance rule.

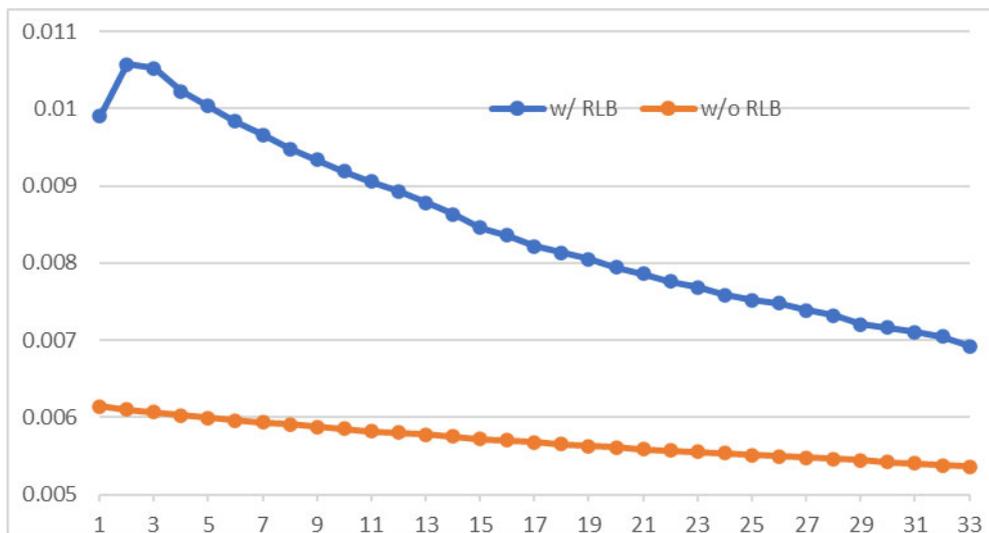

**Figure 6.** Values of $D(|\bigotimes_{i=1}^{5} S_i|)/|\bigotimes_{i=1}^{5} S_i|$ for ID = 1, 2, …, 33.

Figure 6 reveals the affections of the dominance rule, and both ratios are less than 0.01057853 and 0.006140 for values of $D(|\bigotimes_{i=1}^{5} S_i|)/|\bigotimes_{i=1}^{5} S_i|$ with $R_{LB}$ and without $R_{LB}$, respectively. Hence, the proposed dominance rule can reduce the number of solutions up to 99.386% and 98.942% with $R_{LB}$ and without $R_{LB}$, respectively.

Both lines are deceased from 0.00990 to 0.00692 and from 0.00614 to 0.00536 linearly with $R_{LB}$ and without $R_{LB}$, respectively. Hence, the larger the ID the more number of solutions are saved after using the dominance rule.

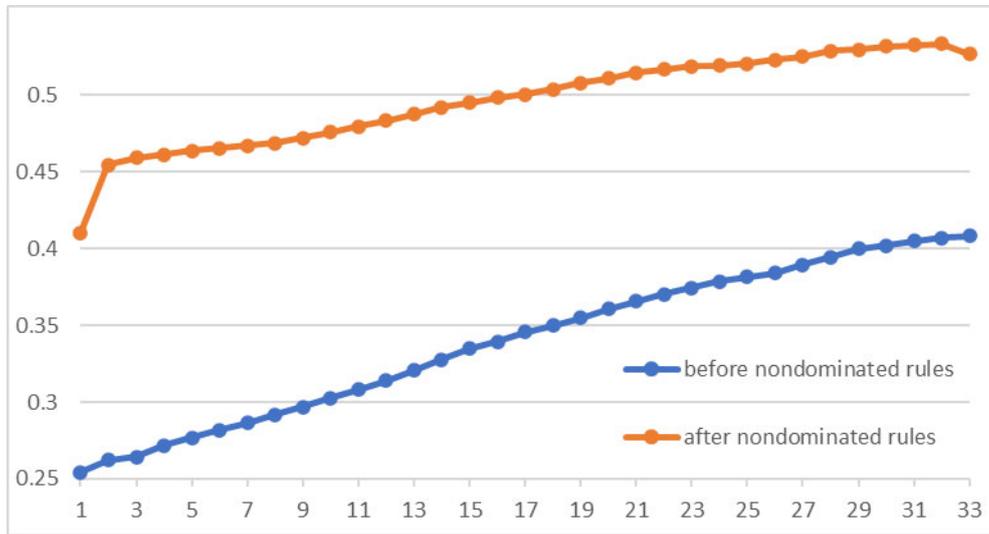

**Figure 7.** The values of $(|\bigotimes_{i=1}^{6} S_i|$ with $R_{LB})/(|\bigotimes_{i=1}^{6} S_i|$ without $R_{LB})$ for ID = 1, 2, …, 33.

In Figure 7, both lines are deceased from 0.00990 to 0.00692 and from 0.00614 to 0.00536 linearly before and after the dominance rule, respectively. Hence, the larger the ID, the more number of solutions are saved after using the dominance rule.

Both ratios are less than 0.408056324 and 0.526596797 before and after using the dominance rule. The number of solutions is reduced to at most 52.66% to their original after using the $R_{LB}$. Such reduction is less than that in Figure 6. Hence, the dominance rule can improve more than that of $R_{LB}$, that is, the dominance rule is more important than $R_{LB}$.

**5.5 Runtime for the BRB**

Each $T_R$ of the proposed BRB is less than 10 seconds which is far less than the runtime $T$ of SSO [20]. For example, in ID 1, $T_R = 1.283$ and $T = 16.5$ for the BRB and SSO, respectively. Moreover, each reliability obtained from the BRB is optimum and that of the SSO is unsure whether it is an optimum or only a near-optimum before we provide the real optimum.

The runtime of the proposed BRB without $R_{UB}$ is also shown in $T$ of Table 6 for each ID. From Table 6, the ratios of $T/T_R$ are decreased from 12.88 at ID = 1 down to 6.48 at ID = 33. Hence, $R_{LB}$ can improve the efficiency more for minor problems than larger-scale problems.

Note that $R_{LB}$ can be the reliability of a solution obtained from any algorithm and not limited to be that of SSO proposed in [20]. Thus, the proposed BRB is very efficient.

**6. CONCLUSIONS & FUTURE WORKS**

In this work, a new BAT called the BRB integrated the upper-bound BAT, dynamic bounds, the dominance rule, and the multiplication of the super sub-BATs are proposed to solve the exact solutions of the 33-variation Fyffe RAP (with mixed components).

To the author's best knowledge, there are no existing algorithms that can solve the exact solutions of the Fyffe RAP while providing the real runtimes. The proposed BRB can have correct solutions and more efficiency if $R_{ub}$ is adapted from the SSO than the metaheuristic-based algorithms, that is, SSO, ACO, GA, and ABC.

The superiority of the proposed BRB has been demonstrated to solve the 33-variation Fyffe RAP systematically and efficiently. The proposed BRB will be extended and enhanced to solve real engineering optimization problems with more variables to strengthen BRB performance in future research.

ACKNOWLEDGEMENT

This research was supported in part by the National Science Council of Taiwan, R.O.C. under grant MOST 110-2221-E-007-107-MY3.

- 36 -